\documentclass[amsmath,amssymb,onecolumn,draftcls]{IEEEtran}

 \usepackage{color}

\usepackage{graphicx}
\usepackage{amsfonts,amsmath,amssymb}
\usepackage{mathrsfs}
\newtheorem{theorem}            {Theorem}[section]
\newtheorem{corollary}          [theorem]{Corollary}

\newtheorem{lemma}              [theorem]{Lemma}
\newtheorem{sideremark}         [theorem]{Remark}
\newtheorem{sideeg}           [theorem]{Example}
\newtheorem{sideconj}           [theorem]{Conjecture}
\newtheorem{sideassumption}   [theorem]{Assumption}

\newenvironment{remark}         {\begin{sideremark}\rm}{\end{sideremark}}
\newenvironment{example}         {\begin{sideeg}\rm}{\end{sideeg}}

\usepackage{enumerate,cite,latexsym,graphicx}
\begin{document}

\title{Quantum Dissipative Systems and Feedback Control  Design by Interconnection}
\author{Matthew R.~James,\thanks{M.R. James is with the Department of
    Engineering, Australian
    National University, Canberra, ACT 0200,
    Australia. Matthew.James@anu.edu.au. Research supported by the
    Australian Research Council.}
    \and
     John~Gough\thanks{J.~Gough is with the Institute for Mathematical and Physical Sciences, University of Wales,
Aberystwyth, Ceredigion, SY23 3BZ, Wales. jug@aber.ac.uk}
     }

\date{\today}

\maketitle

%\tableofcontents

%%%%%%%%%%%%%%%%%%%%%%%%%%%%%%%%%%%%%%%%%%%%%%%%%%%%%%%%%%%%%%%%%%%%%%%%%%%%%%%%
\begin{abstract}
The purpose of this paper is to extend J.C.~Willems' theory of
dissipative systems to the quantum domain. This general theory,
which combines  perspectives from the quantum physics and control
engineering communities, provides useful methods for analysis and
design of dissipative quantum systems. We describe the interaction
of the plant and a class of exosystems in general quantum feedback
network terms. Our results include an infinitesimal
characterization of the dissipation property, which generalizes
the well-known Positive Real and Bounded Real Lemmas, and is used
to study some properties of quantum dissipative systems. We also
show how to formulate control design problems using quantum
network models, which implements Willems' \lq\lq{control by
interconnection}\rq\rq \   for open quantum systems. This control
design formulation includes, for example, standard problems of
stabilization, regulation, and robust control.

\end{abstract}

%%%%%%%%%%%%%%%%%%%%%%%%%%%%%%%%%%%%%%%%%%%%%%%%%%%%%%%%%%%%%%%%%%%%%%%%%%%%%%%%

\section{Introduction}
\label{sec:introduction}

\begin{figure}[h]
\begin{center}
\begin{picture}(0,0)%
\includegraphics{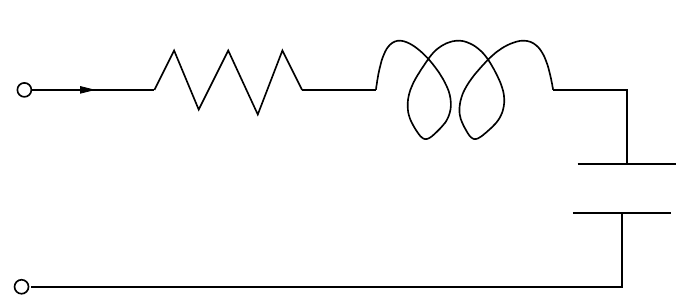}%
\end{picture}%
\setlength{\unitlength}{2072sp}%
\begingroup\makeatletter\ifx\SetFigFont\undefined%
\gdef\SetFigFont#1#2#3#4#5{%
  \reset@font\fontsize{#1}{#2pt}%
  \fontfamily{#3}\fontseries{#4}\fontshape{#5}%
  \selectfont}%
\fi\endgroup%
\begin{picture}(6285,2691)(4441,-6664)
\put(5626,-4156){\makebox(0,0)[lb]{\smash{{\SetFigFont{6}{7.2}{\familydefault}{\mddefault}{\updefault}{\color[rgb]{0,0,0}$+$}%
}}}}
\put(5041,-4651){\makebox(0,0)[lb]{\smash{{\SetFigFont{6}{7.2}{\familydefault}{\mddefault}{\updefault}{\color[rgb]{0,0,0}$i$}%
}}}}
\put(4456,-6361){\makebox(0,0)[lb]{\smash{{\SetFigFont{6}{7.2}{\familydefault}{\mddefault}{\updefault}{\color[rgb]{0,0,0}$-$}%
}}}}
\put(6121,-5326){\makebox(0,0)[lb]{\smash{{\SetFigFont{6}{7.2}{\familydefault}{\mddefault}{\updefault}{\color[rgb]{0,0,0}$R$}%
}}}}
\put(4456,-5776){\makebox(0,0)[lb]{\smash{{\SetFigFont{6}{7.2}{\familydefault}{\mddefault}{\updefault}{\color[rgb]{0,0,0}$v$}%
}}}}
\put(4456,-5101){\makebox(0,0)[lb]{\smash{{\SetFigFont{6}{7.2}{\familydefault}{\mddefault}{\updefault}{\color[rgb]{0,0,0}$+$}%
}}}}
\put(10666,-5191){\makebox(0,0)[lb]{\smash{{\SetFigFont{6}{7.2}{\familydefault}{\mddefault}{\updefault}{\color[rgb]{0,0,0}$+$}%
}}}}
\put(10666,-5731){\makebox(0,0)[lb]{\smash{{\SetFigFont{6}{7.2}{\familydefault}{\mddefault}{\updefault}{\color[rgb]{0,0,0}$v_C$}%
}}}}
\put(9361,-5776){\makebox(0,0)[lb]{\smash{{\SetFigFont{6}{7.2}{\familydefault}{\mddefault}{\updefault}{\color[rgb]{0,0,0}$C$}%
}}}}
\put(8371,-5551){\makebox(0,0)[lb]{\smash{{\SetFigFont{6}{7.2}{\familydefault}{\mddefault}{\updefault}{\color[rgb]{0,0,0}$L$}%
}}}}
\put(10711,-6226){\makebox(0,0)[lb]{\smash{{\SetFigFont{6}{7.2}{\familydefault}{\mddefault}{\updefault}{\color[rgb]{0,0,0}$-$}%
}}}}
\put(9631,-4201){\makebox(0,0)[lb]{\smash{{\SetFigFont{6}{7.2}{\familydefault}{\mddefault}{\updefault}{\color[rgb]{0,0,0}$-$}%
}}}}
\put(8371,-4156){\makebox(0,0)[lb]{\smash{{\SetFigFont{6}{7.2}{\familydefault}{\mddefault}{\updefault}{\color[rgb]{0,0,0}$v_L$}%
}}}}
\put(7741,-4156){\makebox(0,0)[lb]{\smash{{\SetFigFont{6}{7.2}{\familydefault}{\mddefault}{\updefault}{\color[rgb]{0,0,0}$+$}%
}}}}
\put(6976,-4156){\makebox(0,0)[lb]{\smash{{\SetFigFont{6}{7.2}{\familydefault}{\mddefault}{\updefault}{\color[rgb]{0,0,0}$-$}%
}}}}
\put(6166,-4156){\makebox(0,0)[lb]{\smash{{\SetFigFont{6}{7.2}{\familydefault}{\mddefault}{\updefault}{\color[rgb]{0,0,0}$v_R$}%
}}}}
\end{picture}%

\caption{Series RLC circuit.}
\label{fig:rc1}
\end{center}
\end{figure}
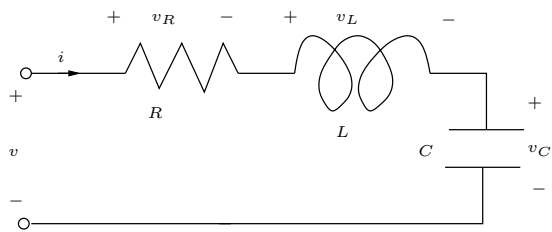

In 1972 J.C.~Willems \cite{WIL72}  developed a general theory of
dissipative systems for the purpose of stability analysis of open
systems, that is, systems that may be subject to external
influences, \cite{JW98}. This theory generalizes Lyapunov methods
that apply to closed systems, as well as  important  results in
control  theory including the positive and bounded real lemmas.
The theory is widely used in control system analysis and design.
In particular, methods for stability analysis, e.g. \cite{PM74},
\cite{HM76}, \cite{HM80},  control design by energy shaping,
interconnection,  robust control system design;  e.g. \cite{OS89},
\cite{VDS96}, \cite{JW97}, \cite{JW98},   \cite{OSME02},
\cite{JW03} have been developed.  While Willems' theory applies to
quite general nonlinear systems, it has its origins in classical
physical systems and is based on describing   energy storage and
flows. As a simple example, consider the passive series RLC circuit of
Figure \ref{fig:rc1}.   Energy  $V= \frac{L}{2} i^2 + \frac{1}{2C}
q^2$ is stored in the inductor and the capacitor (here, $L$
is the impedance,  $C$ is the capacitance, $q$ is capacitor charge, and $i$ is the current). The external voltage
$v$ is related to the internal voltages by $v = v_L +v_R +v_C$,
with $v_L = L \ddot{q}$, $v_R = R \dot{q}$ and $v_C =\frac{1}{C}
q$, and this determines the dynamical equation $\frac{di}{dt}
=\frac{1}{L} (v -Ri-\frac{1}{C}q)$. The rate of change of stored
energy is given by
\begin{equation}
\dot V = - R i^2 + vi .
\label{rc-diss-1}
\end{equation}
The first term on the RHS is the energy dissipated by the
resistor, and the second term is the power applied to the circuit
by an external source. This expression
describes the balance of energy flows, and implies the inequality
\begin{equation}
\dot V \leq   vi ,
\label{rc-diss-2}
\end{equation}
which is an example (in differential form) of the {\em dissipation
inequalities} considered by Willems. It says simply that the rate
at which energy is stored must be less than the rate at which
energy is supplied; the remainder being  dissipated. Inequalities
of this type are of fundamental importance to stability analysis
of open systems.  Furthermore, such inequalities can be exploited
to facilitate control system design. Because of the emerging need
for analysis and design methods for  quantum technologies, the
purpose of this paper is to develop an analogous framework for
open quantum systems.

\begin{figure}[h]
\begin{center}

 \begin{picture}(0,0)%
\includegraphics{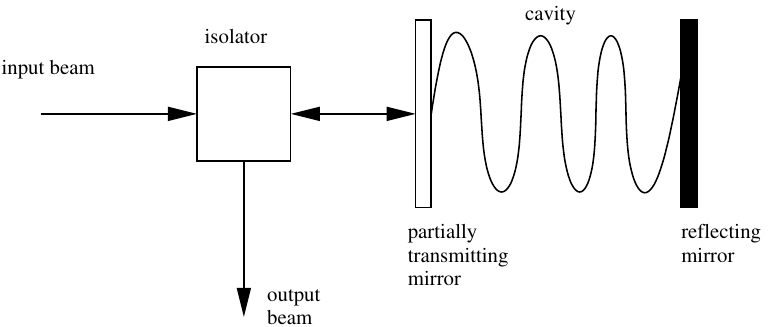}%
\end{picture}%
\setlength{\unitlength}{1973sp}%
\begingroup\makeatletter\ifx\SetFigFont\undefined%
\gdef\SetFigFont#1#2#3#4#5{%
  \reset@font\fontsize{#1}{#2pt}%
  \fontfamily{#3}\fontseries{#4}\fontshape{#5}%
  \selectfont}%
\fi\endgroup%
\begin{picture}(7385,3111)(-6689,-5401)
\put(-6524,-3736){\makebox(0,0)[lb]{\smash{{\SetFigFont{6}{7.2}{\familydefault}{\mddefault}{\updefault}{\color[rgb]{0,0,0}$B$}%
}}}}
\put(-5174,-4861){\makebox(0,0)[lb]{\smash{{\SetFigFont{6}{7.2}{\familydefault}{\mddefault}{\updefault}{\color[rgb]{0,0,0}$\tilde{B}$}%
}}}}
\end{picture}%

\caption{A  cavity consists of a pair of mirrors, one of which is
perfectly reflecting (shown solid) while the other is  partially
transmitting (shown unfilled). The partially transmitting mirror
enables the light mode inside the cavity to interact with an
external light field, such as a laser beam. The external field is
separated into input and output components by a Faraday isolator.}
\label{fig:cavity}
\end{center}
\end{figure}
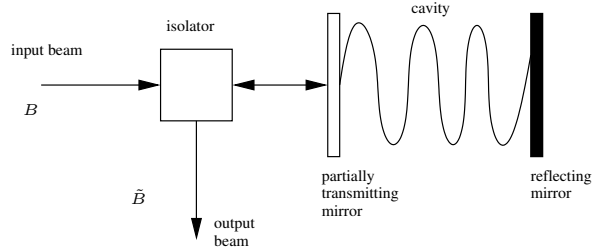

The types of open quantum systems we consider include the optical
cavity shown in Figure \ref{fig:cavity}, which consists of a pair
of mirrors (one of which is partially transmitting) supporting a
trapped mode of light. This cavity mode  may interact with a free
external optical field through the partially transmitting mirror.
The external field consists of two components: the input field,
which is the field before it has interacted with the cavity mode,
and the output field, being the field after interaction. The
output field may carry away energy, and in this way the cavity
system dissipates energy. This quantum system is in some ways
analogous to the RLC circuit discussed above, which stores
electromagnetic energy in the inductor and capacitor, but loses energy as
heat through the resistor. The cavity also stores electromagnetic energy, quantized as 
photons, and these  may be lost to the external field. If $V$
denotes the physical observable for the energy of the cavity mode,
and if a laser beam is applied as an input $w$, then the energy
dissipation relation for the cavity is of the form
\begin{equation}
\mathcal{G}(V) \leq z^\ast w + w^\ast z ,
\label{cavity-diss-1}
\end{equation}
where $z$ is a physical  quantity of the cavity (notation is
explained at the end of this section). The term $\mathcal{G}(V)$
plays the role of $\dot V$, and is discussed further in sections
\ref{sec:open-prelim} and \ref{sec:open-defs}. This inequality
relates the rate at which energy is stored in the cavity to the
rate at which energy is supplied, with the remainder being lost to
the external  field (which serves as a heat bath).

Our principal goal in this paper is to formalize a notion of
dissipation for open quantum systems in a way that is helpful for
quantum control analysis and design by  combining  perspectives
from control engineering with perspectives from quantum physics.
Indeed, phenomena of {\em dissipation} or {\em damping} are
fundamental in both physics and engineering, and have been the
subject of extensive investigations. In physics, methods have been
developed to model energy loss and decoherence (loss of quantum
coherence) arising from the interaction of a system with an
environment, or heat bath, \cite{HN28}, \cite{YD84}, \cite{HP84},
\cite{GC85}, \cite{HC93}, \cite{GZ00}. Open quantum models are
unitary models that provide a description of the system of
interest, or {\em plant}  (e.g. optical cavity, atom, etc) as well
as the environment (e.g. optical field), where the  influence of
the environment on the system may be described with the aid of
{\em quantum noise}. These models preserve the essential nature of
quantum mechanics, a feature of basic importance.

Our modeling of dissipative quantum systems has its origins in the
papers \cite{FJ96}, \cite{DJP00}, \cite{DJ06}, \cite{JNP08}, and
represents inputs as the sum of a quantum noise term, used in
physics  to describe damping as mentioned above, and a signal term
of the type used by Willems to provide a means for the external
world to influence the system. This signal plus noise
representation arises in a natural way in  the  unitary models  we
use.  The signal component of the input comes from the output of
another system, which we call  an {\em exosystem}. We develop our
framework for quantum dissipative systems by using open quantum
model descriptions of {\em quantum networks},  \cite{GJ08},
\cite{GJ08a}, \cite{CWG93}, \cite{HJC93}, \cite{YK03a},
\cite{YK03b}. We therefore employ a network description of a plant
interacting with an exosystem. We allow the exosystem to vary in a
class of such exosystems, and the dissipation inequality we define
is expressed in these terms. This network description turns out to
be very appropriate and efficient, and was inspired  by the
behavioral  \lq\lq{control as interconnection}\rq\rq\  perspective
in Willems' more recent work, \cite{JW97}, \cite{JW98} (also
\cite{TW99}, \cite{TW02}, \cite{WT02}). We provide an
infinitesimal characterization of the dissipation property, and
this important tool is used indispensably in this paper. This
characterization includes generalizations of the well-known
Positive Real and Bounded Real Lemmas, and is used to study some
properties of quantum dissipative systems. We show that
essentially all open quantum systems are dissipative for a
suitable choice of supply rate. This \lq\lq{natural}\rq\rq \
supply rate includes terms corresponding to the classical notion
of {\em passivity}, a dissipation or damping term due to the
quantum noise, and a dissipative term which is present in general
when the exosystem does not commute with the plant.

The framework we  develop can also be used to describe how systems
are influenced by controllers, and hence is useful for control
design by interconnection,  \cite{JW97}, \cite{JW98}, and various
passivity-based methods \cite{OS89},   \cite{VDS96},
\cite{OSMM01}, \cite{OSME02}. We give a general description of how
to formulate control design problems in these terms. This
description includes, for example,  standard problems of
stabilization, regulation, and robust control.  It is important to
appreciate that because we express control design problems in
terms of unitary models, the controllers obtained can (in
principle) be physically realized---this is vital when we wish,
for instance, the controller to be itself a quantum system, as in
{\em coherent control}, \cite{SL00}, \cite{WM94b}, \cite{JNP08},
\cite{NJP08}, \cite{NJD08}, \cite{HM08}.

Note that issues of stability are important in the analysis and
design of quantum networks. This is because quantum networks may
contain active elements that introduce energy, and whenever this
happens, stability of feedback loops in the network is a basic
consideration (small gain theorem, \cite{GZ66a}, \cite{ZDG96},
\cite{DJ06}). Energy may be introduced by design, such as via an
amplifier, or by accident due to  undesirable environmental
influences. It also plays a role in regulation of a system to a
desired equilibrium mode of operation.

We begin in section \ref{sec:open}   by describing the
mathematical models for quantum feedback networks we use, which
are expressed in terms of  the quantum stochastic calculus,
\cite{HP84}, \cite{GC85}. This section includes some material
aimed at helping orient the reader to the ideas, models and
notation used in the remainder of this paper.
Our main definitions and results for quantum dissipative systems
are given in section \ref{sec:qdiss}, which includes some examples
for illustration. Section \ref{sec:qcontrol} contains a
formulation of control by interconnection methodology for quantum
dissipative systems, which we illustrate using simple examples.
The appendix contains some  definitions and results needed in the
paper.

{\em Background references.} A number of articles and books  are
available to help readers with the background material on which
the present paper is based. The papers \cite{YK03a} and
\cite{HSM05} provide excellent introductions to aspects of the
quantum models we use. The paper \cite{BHJ07} is a tutorial
article written to assist control theorists and engineers by
providing introductory discussions of quantum mechanics, open
quantum stochastic models, and quantum filtering. The book
\cite{GZ00} is an invaluable resource for quantum noise models and
quantum optics, while the book \cite{KRP92} provides a detailed
mathematical treatment of the Hudson-Parthasarathy theory of the
quantum stochastic calculus. The book \cite{EM98} is a standard
textbook on quantum mechanics. The papers \cite{GJ08} and
\cite{GJ08a} contain the basic results concerning quantum feedback
networks used in this paper.

{\em Notation.} In this paper we use matrices $ {M} = \{ m_{ij}
\}$ with entries $m_{ij}$ that are operators on an underlying
Hilbert space. The asterisk $\ast$ is used to indicate the Hilbert
space adjoint $A^\ast$ of an operator $A$, as well as  the complex
conjugate $z^\ast = x-iy$ of a complex number $z=x+i y$ (here,
$i=\sqrt{-1}$ and $x,y$ are real).  Real and imaginary parts are
denoted $\mathrm{Re}(z)=(z+z^\ast)/2$ and
$\mathrm{Im}(z)=-i(z-z^\ast)/2$ respectively. The conjugate
transpose $ {M}^\dagger$ of a matrix $ {M}$ is defined by $
{M}^\dagger = \{ m_{ji}^\ast \}$. Also defined are the conjugate $
{M}^\ast = \{ m_{ij}^\ast \}$ and transpose $ {M}^T = \{ m_{ji}
\}$ matrices, so that $ {M}^\dagger=( {M}^T)^\ast=( {M}^\ast)^T$.
In the physics literature, it is common to use the dagger
$\dagger$ to indicate the Hilbert space adjoint. The commutator of
two operators $A,B$ is defined by $[A,B]=AB-BA$. $\delta(\cdot)$
is the Dirac delta function, and $\delta_{jk}$ is the Kronecker
delta. The tensor product  of operators $A$, $B$ defined on
Hilbert spaces $\mathsf{H}$, $\mathsf{G}$ is an operator $A
\otimes B$ defined on the Hilbert space $\mathsf{H} \otimes
\mathsf{G}$ (tensor product of Hilbert spaces) defined by $(A
\otimes B) (\psi \otimes \phi) = (A\psi)  \otimes (B\phi)$ for
$\psi \in \mathsf{H}$, $\phi \in \mathsf{G}$; we usually follow
the standard shorthand and write simply $AB = A\otimes B$ for the
tensor product, and also $A=A\otimes I$ and $B=I \otimes B$.

\section{Preliminaries}
\label{sec:open}

\subsection{The Classical RLC Circuit Revisited}
\label{sec:open-prelim}

Before embarking on a description of the class of open quantum
systems considered in this paper, we revisit the simple classical
RLC circuit of Figure \ref{fig:rc1} in order to provide some
conceptual and notational connections with the quantum theory. We
may choose the charge and current as the \lq\lq{state
variables}\rq\rq \ for the circuit. That is $x=(q,i)^T$ give global
coordinates for the state space manifold which is here a
phase-plane, and the trajectories are then the solutions to the
system of equations $\dfrac{dq}{dt} = w_v^q (q,i),\,
\dfrac{di}{dt} = w_v^i (q,i)$, where $w_v^q (q,i)=i$ and $w_v^i
(q,i) = \frac{1}{L} (v -Ri -\frac{1}{C}q )$. Alternatively, the
trajectories $x(t)=(q(t),i(t))^T$ can be thought of as the integral curves to the
tangent vector field
\begin{equation}
w_v(q,i)   = w_v^q (q,i) \frac{\partial}{\partial q} + w_v^i (q,i)
\frac{\partial}{\partial i}; \label{rlc-x}
\end{equation}
i.e. solutions to the differential equation
\begin{equation}
 \dot x = w_v(x) .
  \label{rlc-xx}
\end{equation}
Note that we need to prescribe the value of the external voltage
$v$, possibly as a function of time, and that $w_v$ actually
corresponds to a family of tangent vector fields generating a
controlled flow.

For an arbitrary smooth function $f(x)=f(q,i)$ of the state, we have by the chain rule
\begin{equation}
\frac{d}{dt} f( x(t) ) = w_v^q (x(t) ) \frac{\partial f}{\partial q}(x(t))  + w_v^i (x(t) )\frac{\partial f}{\partial i} (x(t)),
\label{rlc-f}
\end{equation}
or in compact form
\begin{equation}
\dot f = \mathcal{G}_v (f)
\label{rlc-ff}
\end{equation}
where $\mathcal{G}_v (f)$ is the directional derivative operator
\begin{equation}
 \mathcal{G}_v (f) =   \nabla f  \cdot w_v .
\label{rlc-fff}
\end{equation}
 In particular, the stored energy $V=\frac{L}{2} i^2 + \frac{1}{2C}
q^2$ is a quadratic function of the state variables, and the
energy balance equation (\ref{rc-diss-1}) is obtained from setting
$f=V$:
\begin{equation}
\dot V = \mathcal{G}_v(V) \equiv r^0(v),
 \label{rc-phi-2}
\end{equation}
where $r^0(v)$ is the function $-Ri^2 +iv$ on the phase plane, also
dependent on the input voltage $v$.

The energy balance relation (\ref{rc-diss-1}) and dissipation
inequality (\ref{rc-diss-2}) become
\begin{equation}
\mathcal{G}_v(V) -r^0(v) = 0 , 
\label{rc-diss-1-a}
\end{equation}
and
\begin{equation}
\mathcal{G}_v(V) -r(v) \leq  0 , 
\label{rc-diss-2-a}
\end{equation}
where the respective supply rates are $r^0(v) = -Ri^2 +iv$ and $r(v)
= iv$. If we regard the voltage $v$ as an input and the current
$i$ as an output function of the phase space  variables, then it can be seen that the supply rates are
functions of the phase space variables and the input.

Functions $f=f(q,i)$ of the phase space variables correspond to the
physical variables of interest, and are basic to any description
of classical mechanics. When considering open physical systems,
such as the RLC circuit, or the open quantum systems discussed in
this paper, it is helpful to have a notation for indicating to
which physical system a physical variable belongs. If we write
$\mathscr{A}_{RLC}$ for RLC circuit physical variables (the set
of smooth functions on the phase space manifold) then this is in
fact a commutative algebra which may be extended to a
$\ast$-algebra by taking complex-valued functions with the choice
of complex conjugation as the $\ast$-operation. Likewise we could
write $\mathscr{A}_{ex}$ for functions of the external variables
$v$. We then see that the energy function $V$ belongs to
$\mathscr{A}_{RLC}$. The supply rates however belong to the
algebra of functions over both state variables and external
variables,  the tensor product
$\mathscr{A}_{RLC} \otimes \mathscr{A}_{ex}$, so that $r^0, r \in \mathscr{A}_{RLC} \otimes \mathscr{A}_{ex}$.
In the quantum setting, these algebras will be
non-commutative (in fact algebras of operators over Hilbert
spaces).

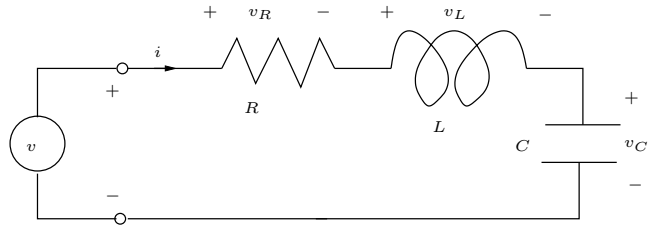
\begin{figure}[h]
\begin{center}
\begin{picture}(0,0)%
\includegraphics{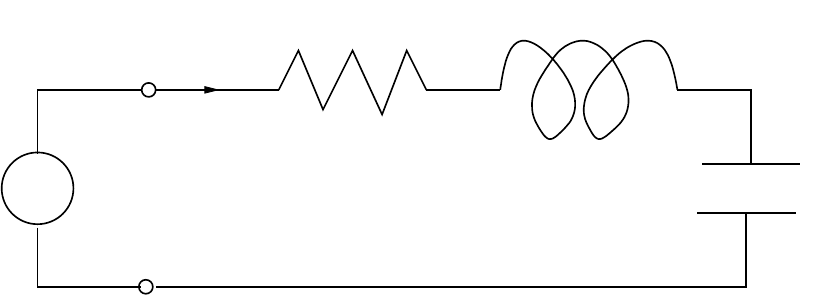}%
\end{picture}%
\setlength{\unitlength}{2072sp}%
\begingroup\makeatletter\ifx\SetFigFont\undefined%
\gdef\SetFigFont#1#2#3#4#5{%
  \reset@font\fontsize{#1}{#2pt}%
  \fontfamily{#3}\fontseries{#4}\fontshape{#5}%
  \selectfont}%
\fi\endgroup%
\begin{picture}(7423,2691)(3303,-6664)
\put(3511,-5776){\makebox(0,0)[lb]{\smash{{\SetFigFont{6}{7.2}{\familydefault}{\mddefault}{\updefault}{\color[rgb]{0,0,0}$v$}%
}}}}
\put(5041,-4651){\makebox(0,0)[lb]{\smash{{\SetFigFont{6}{7.2}{\familydefault}{\mddefault}{\updefault}{\color[rgb]{0,0,0}$i$}%
}}}}
\put(4456,-6361){\makebox(0,0)[lb]{\smash{{\SetFigFont{6}{7.2}{\familydefault}{\mddefault}{\updefault}{\color[rgb]{0,0,0}$-$}%
}}}}
\put(6121,-5326){\makebox(0,0)[lb]{\smash{{\SetFigFont{6}{7.2}{\familydefault}{\mddefault}{\updefault}{\color[rgb]{0,0,0}$R$}%
}}}}
\put(4456,-5101){\makebox(0,0)[lb]{\smash{{\SetFigFont{6}{7.2}{\familydefault}{\mddefault}{\updefault}{\color[rgb]{0,0,0}$+$}%
}}}}
\put(10666,-5191){\makebox(0,0)[lb]{\smash{{\SetFigFont{6}{7.2}{\familydefault}{\mddefault}{\updefault}{\color[rgb]{0,0,0}$+$}%
}}}}
\put(10666,-5731){\makebox(0,0)[lb]{\smash{{\SetFigFont{6}{7.2}{\familydefault}{\mddefault}{\updefault}{\color[rgb]{0,0,0}$v_C$}%
}}}}
\put(9361,-5776){\makebox(0,0)[lb]{\smash{{\SetFigFont{6}{7.2}{\familydefault}{\mddefault}{\updefault}{\color[rgb]{0,0,0}$C$}%
}}}}
\put(8371,-5551){\makebox(0,0)[lb]{\smash{{\SetFigFont{6}{7.2}{\familydefault}{\mddefault}{\updefault}{\color[rgb]{0,0,0}$L$}%
}}}}
\put(10711,-6226){\makebox(0,0)[lb]{\smash{{\SetFigFont{6}{7.2}{\familydefault}{\mddefault}{\updefault}{\color[rgb]{0,0,0}$-$}%
}}}}
\put(9631,-4201){\makebox(0,0)[lb]{\smash{{\SetFigFont{6}{7.2}{\familydefault}{\mddefault}{\updefault}{\color[rgb]{0,0,0}$-$}%
}}}}
\put(7741,-4156){\makebox(0,0)[lb]{\smash{{\SetFigFont{6}{7.2}{\familydefault}{\mddefault}{\updefault}{\color[rgb]{0,0,0}$+$}%
}}}}
\put(6976,-4156){\makebox(0,0)[lb]{\smash{{\SetFigFont{6}{7.2}{\familydefault}{\mddefault}{\updefault}{\color[rgb]{0,0,0}$-$}%
}}}}
\put(6166,-4156){\makebox(0,0)[lb]{\smash{{\SetFigFont{6}{7.2}{\familydefault}{\mddefault}{\updefault}{\color[rgb]{0,0,0}$v_R$}%
}}}}
\put(5626,-4156){\makebox(0,0)[lb]{\smash{{\SetFigFont{6}{7.2}{\familydefault}{\mddefault}{\updefault}{\color[rgb]{0,0,0}$+$}%
}}}}
\put(8506,-4156){\makebox(0,0)[lb]{\smash{{\SetFigFont{6}{7.2}{\familydefault}{\mddefault}{\updefault}{\color[rgb]{0,0,0}$v_L$}%
}}}}
\end{picture}%

\caption{Series RLC circuit connected to an external signal generator.}
\label{fig:rc3}
\end{center}
\end{figure}

The crucial observation is that the energy balance relation
(\ref{rc-diss-1}) and the dissipation inequality (\ref{rc-diss-2})
are to hold for all input signals $v$. This can be re-expressed as
saying that (\ref{rc-diss-1}) and (\ref{rc-diss-2}) hold for all
external signal generators $v$, as shown in Figure \ref{fig:rc3}.
This leads to an interpretation  of energy balance and dissipation
inequalities in terms of an interconnected system consisting of
the principal system of interest (the RLC circuit) and an
exosystem (the signal generator). This interpretation will be used
in the quantum context in section \ref{sec:qdiss}  below.

\subsection{Quantum Mechanics}
\label{sec:open-qm}

In quantum mechanics \cite{EM98} physical quantities like energy,
spin, position, etc., are expressed as {\em observables}; these
are represented as self-adjoint operators acting on a Hilbert
space $\mathsf{H}$. Other physical variables, like annihilation
operators (see Appendix \ref{sec:app-qho})---which are not
self-adjoint---are also of importance.  We will use the notation
$\mathscr{A}$ to refer to the collection of physical variables for
a system (in general $\mathscr{A}$ is a non-commutative
$\ast$-algebra).  We refer to $\mathscr{A}$ as the {\em physical variable space} for the system.
Unit vectors $\psi \in \mathsf{H}$ are called
{\em state vectors}. When a quantum system is in a state defined
by a state vector $\psi \in \mathsf{H}$, the expected value of an
observable $A \in \mathscr{A}$  is defined in terms of the Hilbert
space inner product: $\langle \psi, A \psi \rangle$. In what
follows we use the shorthand notation $\langle A \rangle$ to
denote expectation when the underlying state is understood.

The postulates of quantum mechanics state that for a closed system
the evolution of states and observables are given in terms of a
unitary operator $U(t)$ satisfying the Schr\"{o}dinger equation
\begin{equation}
\dot U = -i H U,
\label{schrodinger}
\end{equation}
with initial condition $U(0)=I$ (the identity). Here, $H$ is an
observable called the {\em Hamiltonian}, and represents the energy
of the system. State vectors  evolve according to $\psi_t = U(t)
\psi$. Alternatively, we may view state vectors as fixed in time,
while observables are taken to evolve according to $A(t)=U^\ast(t)
A U(t)$: this is the Heisenberg picture. Both pictures are
equivalent and the average of an observable $A$ in state $\psi$ at
time $t$ is given equally by $\langle \psi, A \psi \rangle_t
=\langle \psi_t , A \psi_t \rangle =\langle \psi, A (t) \psi
\rangle $.

In this paper we are interested in open quantum systems - systems
that interact with other systems or an environment. These systems
will be defined in section \ref{sec:open-defs}  in terms of a
stochastic generalization of the Schrodinger equation
(\ref{schrodinger}) involving quantum noise. Before considering
these open system models, we look at  a simple situation of two
interacting systems in the next section.

\subsection{A Pair of Interacting Systems}
\label{sec:open-two}

Consider a pair of independent systems $P$ (the plant) and $W$
(the exosystem, or signal generator). The physical variable spaces
for these systems are denoted $\mathscr{A}_{P}$ and
$\mathscr{A}_{W}$ respectively, and consist of operators defined
on underlying Hilbert spaces $\mathsf{H}_P$ and $\mathsf{H}_W$
respectively. The physical variable space for the combined system
is the tensor product $\mathscr{A}_{P} \otimes \mathscr{A}_{W}$,
consisting of operators on the Hilbert space $\mathsf{H}_P \otimes
\mathsf{H}_W$.  All operators in $\mathscr{A}_{P}$ may be regarded
as operators in $\mathscr{A}_{P} \otimes \mathscr{A}_{W}$ by
identifying $A_P$ with $A_P \otimes I_W$, and similarly for $
\mathscr{A}_{W}$. As a consequence, all variables in
$\mathscr{A}_{P} $ commute with all variables in $\mathscr{A}_{W}
$.

Let $H_P \in \mathscr{A}_{P}$ and $H_W \in \mathscr{A}_{W}$ be the
Hamiltonians for each of the systems, respectively; this would be
enough to specify their dynamics as isolated, closed, systems.
However, we allow them to interact by exchanging energy as
specified by the {\em interaction Hamiltonian}
\begin{equation}
H_{PW}= -i (   K^\ast v - v^\ast K ),
\label{H-int-1}
\end{equation}
where $K \in \mathscr{A}_{P}$ and $v \in \mathscr{A}_{W}$. The
total Hamiltonian for the combined system is $H=H_P+H_W+H_{PW}$,
and the dynamics are given by the Schr\"{o}dinger equation
(\ref{schrodinger}) using this total Hamiltonian.

Now let's  consider the effect of the exosystem $W$ on the plant
$P$. Let $V \in \mathscr{A}_{P}$ be a non-negative observable that
commutes with $H_P$.  Then from (\ref{schrodinger})  we see that
$V$ evolves according to
\begin{eqnarray}
\dot V &=& -i [V, H]
\nonumber
\\
&=& v^\ast [ V, K] - [V, K^\ast ] v
\nonumber
\\
&=& Z^\ast v + v^\ast Z ,
\label{q-passive-easy}
\end{eqnarray}
where $Z=[V, K] \in \mathscr{A}_{P}$. From this we see that $P$
is lossless, with the RHS of (\ref{q-passive-easy}) giving the net
rate at which energy is delivered to $P$  from $W$ (cf.  \cite[eq.
(2.39) and Chapter 4]{VDS96}, and section \ref{sec:qdiss-dfn}
below).

In the general framework we present in this paper (section
\ref{sec:qdiss}), $P$ is an open system, and  $W$  may be
connected to $P$ via field connections in addition to direct
couplings of the form (\ref{H-int-1}). Open quantum systems are
summarized in section \ref{sec:open-defs}, and mechanisms for
interconnecting them are reviewed in section \ref{sec:open-qfn}.

\subsection{Definitions}
\label{sec:open-defs}

We consider an open quantum system $ {G}$ with physical variable
space  $\mathscr{A}_{ {G}}$ consisting of operators $X$ defined on
an underlying Hilbert space $\mathsf{H}_G$. The self-energy of
this system is described by a Hamiltonian $H \in \mathscr{A}_{
{G}}$. This system is driven by a collection of $n$ field channels
given by the quantum stochastic processes
$$
 {A} =  \left( \begin{array}{c}
A_1 \\ \vdots \\ A_n
\end{array} \right),  \ \
 {\Lambda} = \left( \begin{array}{ccc}
A_{11} & \hdots & A_{1n} \\
 \vdots & \vdots & \vdots \\
 A_{n1} & \hdots & A_{nn}
\end{array} \right) .
$$
These  respectively describe annihilation of photons in the field
channels, and  scattering between channels, and are operators on a
Hilbert space $\mathsf{F}$, with associated variable space
$\mathscr{F}$. Specifically, $\mathsf{F}$ is the Hilbert space
describing an indefinite number of quanta (called a Fock space
\cite{KRP92}), and $\mathscr{F}$ is the space of operators over
this space. We assume that these processes are {\em canonical},
meaning that we have the following non-vanishing second order Ito products: $%
dA_{j}\left( t\right) dA_{k}\left( t\right) ^{\ast }=\delta _{jk}dt$, $%
dA_{jk}\left( t\right) dA_{l}\left( t\right) ^{\ast }=\delta
_{kl}dA_{j}(t)^{\ast }$, $\,dA_{j}\left( t\right) dA_{kl}\left(
t\right) =\delta _{jk}dA_{l}(t)$ and $dA_{jk}\left( t\right)
dA_{lm}\left( t\right) =\delta _{kl}dA_{jm}(t)$. The simplest
situation corresponds to that of a vacuum state $\phi \in
\mathsf{F}$ for the field channels, in which case the input
processes are purely quantum noise.

Coupling of the system to the field is defined using
$$
 {S} =   \left( \begin{array}{ccc}
S_{11} & \hdots & S_{1n} \\
 \vdots & \vdots & \vdots \\
 S_{n1} & \hdots & S_{nn}
\end{array} \right) , \ \
 {L} =  \left( \begin{array}{c}
L_1 \\ \vdots \\ L_n
\end{array} \right) ,
$$
respectively a {\em scattering matrix} with operator entries
$S_{ij} \in \mathscr{A}_{{G}}$ satisfying $ {S}^\dagger  {S}=  {S}
{S}^\dagger =  {I}$, and a vector of {\em coupling operators} $L_j
\in \mathscr{A}_{ {G}}$.

In terms of the parameters $ {G}=( {S},  {L}, H)$,
 the Schrodinger equation
\begin{eqnarray}
d U(t) &=&  \left\{ \mathrm{tr}[( {S}- {I}) d  {\Lambda}] + d
{A}^\dagger  {L} - {L}^\dagger  {S} d {A}
%\right.
%\nonumber  \\
%&&  \left.
- \frac{1}{2}  {L}^\dagger  {L} dt -i H dt
\right\} U(t)
\label{open-hp-G}
\end{eqnarray}
with initial condition $U(0)=I$ determines the unitary motion of
the system, in accordance with the fundamental postulate of
quantum mechanics. Given a system operator $X \in \mathscr{A}_{
{G}}$, its Heisenberg evolution is defined by $ X(t) =
\mathsf{j}_t(X)=U\left( t\right) ^{\ast } X
  U\left( t\right)
$ and satisfies
\begin{eqnarray}
dX(t) =  (\mathcal{L}_{ {L}(t) } (X(t)) -i [ X(t), H(t) ])dt
\nonumber \\
 + d {A}^\dagger(t)  {S}^\dagger(t) [ X(t),  {L}(t)]
+ [ {L}^\dagger (t),X(t)]  {S}(t) d {A}(t)
\nonumber \\
+ \mathrm{tr}[ ( {S}^\dagger (t) X(t)  {S}(t) - X(t) ) d {\Lambda}(t)] .
\label{qle-X}
\end{eqnarray}
 In this expression, all operators evolve unitarily  (e.g.
$ {L}(t)=\mathsf{j}_t(  {L})$)  (commutators of vectors and
matrices of operators are defined component-wise), and tr denotes
the trace of a matrix. We also employ the notation
\begin{equation}
\mathcal{L}_{ {L}}(X) =  \frac{1}{2}  {L}^\dagger[X, {L}]
+ \frac{1}{2} [ {L}^\dagger, X ]  {L}  .
\label{lindblad}
\end{equation}
 In what follows we write
\begin{equation}
\mathcal{G}_G(X) = -i [X,H] + \mathcal{L}_L(X)
\label{G-gen-def}
\end{equation}
for the {\em generator} of the plant $G$. The components of the
output fields are defined by $\tilde{ {A}}(t) = \mathsf{j}_t(
{A}(t)) \doteq U^\ast(t)    {A}(t)  U(t)$, $\tilde{ {\Lambda}}(t)
=\mathsf{j}_t( {\Lambda}(t)) \doteq U^\ast(t)    {\Lambda}(t)
U(t)$ and satisfy the quantum stochastic differential equations
\begin{eqnarray}
d\tilde{ {A}}(t) &=&  {S}(t) d {A}(t) +  {L}(t) dt
\label{out-A}  \\
d \tilde { {\Lambda}}(t) &=&    {S}^\ast(t) d {\Lambda}(t)  {S}^T(t) +
   {S}^\ast(t) d {A}^\ast(t)  {L}^T(t)
  \\
  &&+  {L}(t) d {A}(t)  {S}^T(t)
  +  {L}^\ast(t)  {L}^T(t) dt ,
\label{out-Lambda}
\end{eqnarray}
where $ {L}(t) = \mathsf{j}_t( {L})$, etc., as above. The output
processes also have canonical quantum Ito products.

It can be seen that the parameters $ {G} = ( {S},  {L},  H)$
provide a compact specification of the open system, assuming
canonical field inputs, since they determine the behavior of the
system, via the flow $\mathsf{j}_t(\cdot)$, as determined by the
Schrodinger equation (\ref{open-hp-G}).  In the case of a purely
static system, we sometimes use the shorthand $S=(S,0,0)$.
Important special cases are $I=(I,0,0)$, the trivial (identity)
system, and $J=(J,0,0)$, where
\begin{equation*}
J=\left(
\begin{array}{cc} 0 & 1
\\ 1 & 0 \end{array}\right)
.
\end{equation*}

If $\psi \in \mathsf{H}_G$ is an initial system state vector, then
with vacuum fields the state vector  of the complete system is
$\psi \otimes \phi$.  Then the quantum expectation $\langle X(t)
\rangle$ is defined to be $\langle  \psi \otimes \phi, X(t) \psi
\otimes \phi \rangle $. In order to describe how quantum noise
beyond time $t$ is averaged out, we introduce a collection $\{
\mathscr{F}_t \}$ of  physical variable spaces such that
$\mathscr{F}_t \subset \mathscr{A}_{ {G}} \otimes \mathscr{F}$ is
generated by operators in $\mathscr{A}_{ {G}}$ and the quantum
noises $A_{ij}(s)$, $s\leq t$. Then $X(t)$ is adapted, i.e.
$X(t)\in \mathscr{F}_t$, and $\mathscr{F}_0 = \mathscr{A}_{ {G}}$.
There is an associated {\em  vacuum expectation} $\mathbb{E}_t :
\mathscr{A}_{ {G}}   \otimes \mathscr{F} \to \mathscr{F}_t$
\cite[Chapter 26]{KRP92} with respect to which the open dynamics
satisfies
\begin{eqnarray}
\mathbb{E}_s[ X(t) ] = X(s) + \int_s^t \mathbb{E}_s \left[
\mathcal{G}_{ {G}} (X(r))  \right]
  dr
\label{vac-exp}
\end{eqnarray}
for all $t \geq s$. In this expression $\mathbb{E}_s[ X(t) ]$
depends on the initial operators and the quantum noises up to time
$s$, while the noises beyond time $s$ have been averaged out; it
captures the Markovian nature of the model.

\subsection{Quantum Feedback Networks}
\label{sec:open-qfn}

In this section we describe a quantum framework for feedback
networks that will be used in the sequel, \cite{GJ08},
\cite{GJ08a}. {\em Quantum feedback networks}  (QFN) consist of open
quantum components that are interconnected by means of field
channels that serve as \lq\lq{quantum wires}\rq\rq. These channels
enable the directional  transmission of quantum signals, thereby
allowing the components to interact; the components may also
interact directly via suitable couplings that facilitate
bidirectional energy exchanges, as discussed in section
\ref{sec:open-two}. Here we focus on the directional
interconnections. The QFN framework is expressed in terms of
elementary constructs that enable efficient description of
networks. These network constructs are defined in terms of the
open system parameters $(S,L,H)$ discussed in section
\ref{sec:open-defs}. As we will see, the framework generalizes the
familiar transfer function descriptions widely used in classical
linear systems theory; however, we emphasize that the QFN
framework holds for general open quantum components whose
dynamical variables may evolve nonlinearly (by this we mean that
the differential equation for a component operator $X(t)$ may be
nonlinear).

\begin{figure}[h]
\begin{center}

\setlength{\unitlength}{1579sp}%
\begingroup\makeatletter\ifx\SetFigFont\undefined%
\gdef\SetFigFont#1#2#3#4#5{%
  \reset@font\fontsize{#1}{#2pt}%
  \fontfamily{#3}\fontseries{#4}\fontshape{#5}%
  \selectfont}%
\fi\endgroup%
\begin{picture}(4244,4590)(2679,-6514)
\put(4051,-6436){\makebox(0,0)[lb]{\smash{{\SetFigFont{5}{6.0}{\familydefault}{\mddefault}{\updefault}{\color[rgb]{0,0,0}$G_1 \boxplus G_2$}%
}}}}
\thicklines
{\color[rgb]{0,0,0}\put(3901,-5611){\framebox(1800,1500){}}
}%
{\color[rgb]{0,0,0}\put(3901,-4861){\vector(-1, 0){1200}}
}%
\thinlines
{\color[rgb]{0,0,0}\put(3076,-6136){\dashbox{150}(3300,4200){}}
}%
\thicklines
{\color[rgb]{0,0,0}\put(6901,-4861){\vector(-1, 0){1200}}
}%
{\color[rgb]{0,0,0}\put(6901,-3211){\vector(-1, 0){1200}}
}%
{\color[rgb]{0,0,0}\put(3901,-3211){\vector(-1, 0){1200}}
}%
\put(4651,-3286){\makebox(0,0)[lb]{\smash{{\SetFigFont{5}{6.0}{\familydefault}{\mddefault}{\updefault}{\color[rgb]{0,0,0}$G_1$}%
}}}}
\put(4726,-4936){\makebox(0,0)[lb]{\smash{{\SetFigFont{5}{6.0}{\familydefault}{\mddefault}{\updefault}{\color[rgb]{0,0,0}$G_2$}%
}}}}
{\color[rgb]{0,0,0}\put(3901,-3961){\framebox(1800,1500){}}
}%
\end{picture}%

\caption{Concatenation of two systems, $ {G}_1  \boxplus  {G}_2$.}
\label{fig:concat}
\end{center}
\end{figure}
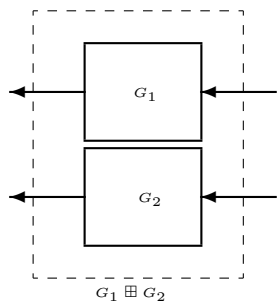

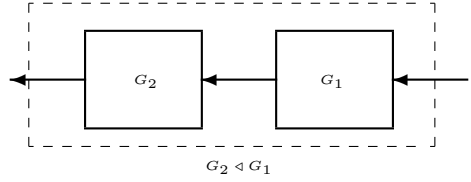
\begin{figure}[h]
\begin{center}

\setlength{\unitlength}{1579sp}%
\begingroup\makeatletter\ifx\SetFigFont\undefined%
\gdef\SetFigFont#1#2#3#4#5{%
  \reset@font\fontsize{#1}{#2pt}%
  \fontfamily{#3}\fontseries{#4}\fontshape{#5}%
  \selectfont}%
\fi\endgroup%
\begin{picture}(7244,2715)(2679,-4714)
\put(5776,-4636){\makebox(0,0)[lb]{\smash{{\SetFigFont{5}{6.0}{\familydefault}{\mddefault}{\updefault}{\color[rgb]{0,0,0}$G_2 \triangleleft G_1$}%
}}}}
\thicklines
{\color[rgb]{0,0,0}\put(3901,-3211){\vector(-1, 0){1200}}
}%
{\color[rgb]{0,0,0}\put(6901,-3961){\framebox(1800,1500){}}
}%
\thinlines
{\color[rgb]{0,0,0}\put(3001,-4261){\dashbox{150}(6375,2250){}}
}%
\thicklines
{\color[rgb]{0,0,0}\put(6901,-3211){\vector(-1, 0){1200}}
}%
{\color[rgb]{0,0,0}\put(9901,-3211){\vector(-1, 0){1200}}
}%
\put(4651,-3286){\makebox(0,0)[lb]{\smash{{\SetFigFont{5}{6.0}{\familydefault}{\mddefault}{\updefault}{\color[rgb]{0,0,0}$G_2$}%
}}}}
\put(7576,-3286){\makebox(0,0)[lb]{\smash{{\SetFigFont{5}{6.0}{\familydefault}{\mddefault}{\updefault}{\color[rgb]{0,0,0}$G_1$}%
}}}}
{\color[rgb]{0,0,0}\put(3901,-3961){\framebox(1800,1500){}}
}%
\end{picture}%

\caption{Series or cascade connection of two systems, $ {G}_2  \triangleleft  {G}_1$.}
\label{fig:series-prod}
\end{center}
\end{figure}

QFN modeling proceeds as follows. Before implementing any
connections, we first collect the components together. This is
described using the concatenation product $\boxplus$, Figure
\ref{fig:concat}. Next, we identify any series connections between
components, which we describe using the series product
$\triangleleft$, Figure \ref{fig:series-prod}.   Networks that can
be completely described using the concatenation and series
products are called {\em reducible networks} (these were studied
in detail in \cite{GJ08}). Any remaining signal connections will
form part of a feedback loop that can be described in terms of a
linear fractional transformation, $F(G)$ \cite{GJ08a}. All direct
couplings between components can be accommodated using an
interaction Hamiltonian of the form (\ref{H-int-1}).

Suppose we are given two such systems:  $G_1=(S_1,L_1,H)$ and
$G_2=(S_2,L_2,H)$, with physical variable spaces
$\mathscr{A}_{G_1}$ and $\mathscr{A}_{G_2}$, respectively. The
products we define below combine these systems to produce new
systems defined  in terms of parameters drawn from the tensor
product of variable spaces $\mathscr{A}_{G_1} \otimes
\mathscr{A}_{G_2}$.

The {\em concatenation} of $G_1$ and $G_2$  is the system $ {G}_1
\boxplus  {G}_2$ defined by
\begin{equation}
 {G}_1  \boxplus {G}_2 = (\left( \begin{array}{cc}
 {S}_1 & 0 \\ 0 & {S}_2 \end{array}\right), \left( \begin{array}{c} {L}_1
 \\ {L}_2\end{array}\right),  H_1+H_2) ,
\label{box-plus-dfn}
\end{equation}
as illustrated in Figure \ref{fig:concat} (where each arrowed line may represent multiple channels).
It is possible to include
zero-dimensional inputs into this scheme as a special case: if a system in
isolation has no inputs then it is a closed dynamical system and  its
dynamics are described by a Hamiltonian $H$. It is convenient just to
write this as $ {G}=\left( \_,\_,H\right) $ with the absence of inputs
denoted by blanks; we then just set $\left( \_,\_,H\right) \boxplus \left(
\_,\_,H^{\prime }\right) :=\left( \_,\_,H+H^{\prime }\right) $ and more
generally $\left( \_,\_,H\right) \boxplus \left( S^{\prime }, {L}%
^{\prime },H^{\prime }\right) =\left( S^{\prime }, {L}^{\prime
},H^{\prime }\right) \boxplus \left( \_,\_,H\right) :=\left( S^{\prime },%
 {L}^{\prime },H+H^{\prime }\right) $.

Now suppose  $G_1=(S_1,L_1,H)$ and $G_2=(S_2,L_2,H)$ have the same
number of field channels (i.e. $L_1$ and $ {L}_2$ have the same
length). Then the {\em series product}   $ {G}_2 \triangleleft
{G}_1$ is defined by
\begin{eqnarray}
 {G}_2 \triangleleft  {G}_1 =  \left(  {S}_2 {S}_1,  {L}_2+ {S}_2 {L}_1,
 H_1+H_2+\mathrm{Im} \{ {L}_2^\dagger  {S}_2  {L}_1 \}   \right),
\label{series-dfn}
\end{eqnarray}
see Figure \ref{fig:series-prod}.

Several useful facts concerning the series product are the
following: (i) given a system $G=(S,L,H)$, we have $G=(I,L,H)
\triangleleft (S,0,0) = (S,0,0) \triangleleft (I, S^\dagger L,
H)$, (ii) the  inverse of a system $G$  exists and is given by
$G^{-1}=(S^\dagger, -S^\dagger L, -H)$, by which it is meant that
$G^{-1} \triangleleft G = G \triangleleft G^{-1} = I=(I,0,0)$, and
(iii) for any two systems $G_1$ and $G_2$ we have $G_2
\triangleleft G_1 = G_1 \triangleleft \tilde{G}_2$ where
$\tilde{G}_2 = G_1^{-1} \triangleleft G_2 \triangleleft G_1
=(S_1^\dagger S_2 S_1, S_1^\dagger(S_2-I)L_1+S_1^\dagger L_2, H_2+
\mathrm{Im} \{     L_2^\dagger (S_2+I)L_1-L_1^\dagger S_2L_1 \}
)$.

For future reference, we mention that the  generators for the
systems formed with the concatenation and series  products are
\begin{eqnarray}
\mathcal{G}_{ {G}_1 \boxplus  {G}_2} (X) &=&  \mathcal{L}_{ {L_1}}
(X) + \mathcal{L}_{ {L_2}} (X)  -i [ X, H_1+H_2 ] = \mathcal{G}_{
{G}_1} (X) + \mathcal{G}_{  {G}_2} (X) ,
\label{lind-gen-concat}  \\
\mathcal{G}_{ {G}_2 \triangleleft  {G}_1} (X) &=&  \mathcal{L}_{ {L_2}+ {S}_2 {L_1}} (X)   -i [ X, H_1+H_2 +\frac{1}{2i}( {L}_2^\dagger  {S}_2  {L}_1 - {L}_1^\dagger  {S}_2^\dagger  {L}_2 )]
\label{lind-gen-series}  \\
&=&  \mathcal{L}_{ {S}_2 {L_1}} (X) + \mathcal{L}_{ {L_2}} (X)
+ {L}_1^\dagger  {S}_2^\dagger [X,  {L}_2] + [ {L}_2^\dagger, X]  {S}_2  {L}_1  -i [ X, H_1+H_2]
\nonumber \\
&=&  \mathcal{L}_{ {L_1}} (X) + \mathcal{L}_{ {L_2}} (X) +  {L}_1^\dagger ( {S}_2^\dagger X  {S}_2 - X) {L}_1
+ {L}_1^\dagger  {S}_2^\dagger [X,  {L}_2] + [ {L}_2^\dagger, X]  {S}_2  {L}_1  -i [ X, H_1+H_2]  .
\nonumber
\end{eqnarray}

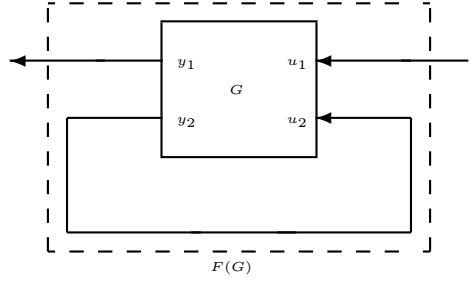
\begin{figure}[h]
\begin{center}

\setlength{\unitlength}{1579sp}%
\begingroup\makeatletter\ifx\SetFigFont\undefined%
\gdef\SetFigFont#1#2#3#4#5{%
  \reset@font\fontsize{#1}{#2pt}%
  \fontfamily{#3}\fontseries{#4}\fontshape{#5}%
  \selectfont}%
\fi\endgroup%
\begin{picture}(7244,4300)(1479,-5539)
\put(4651,-5461){\makebox(0,0)[lb]{\smash{{\SetFigFont{5}{6.0}{\familydefault}{\mddefault}{\updefault}{\color[rgb]{0,0,0}$F(G)$}%
}}}}
\thicklines
{\color[rgb]{0,0,0}\put(7801,-3061){\vector(-1, 0){1500}}
\put(7801,-3061){\line( 0,-1){1800}}
\put(7801,-4861){\line(-1, 0){2100}}
}%
{\color[rgb]{0,0,0}\put(4501,-4861){\line(-1, 0){2100}}
\put(2401,-4861){\line( 0, 1){1800}}
\put(2401,-3061){\line( 1, 0){1500}}
}%
{\color[rgb]{0,0,0}\put(4351,-4861){\line( 1, 0){1650}}
}%
{\color[rgb]{0,0,0}\put(2101,-5161){\dashbox{225}(6000,3900){}}
}%
{\color[rgb]{0,0,0}\put(3001,-2161){\vector(-1, 0){1500}}
}%
{\color[rgb]{0,0,0}\put(7801,-2161){\vector(-1, 0){1500}}
}%
{\color[rgb]{0,0,0}\put(2851,-2161){\line( 1, 0){1050}}
}%
{\color[rgb]{0,0,0}\put(7651,-2161){\line( 1, 0){1050}}
}%
\put(5851,-2236){\makebox(0,0)[lb]{\smash{{\SetFigFont{5}{6.0}{\familydefault}{\mddefault}{\updefault}{\color[rgb]{0,0,0}$u_1$}%
}}}}
\put(4126,-3136){\makebox(0,0)[lb]{\smash{{\SetFigFont{5}{6.0}{\familydefault}{\mddefault}{\updefault}{\color[rgb]{0,0,0}$y_2$}%
}}}}
\put(4126,-2236){\makebox(0,0)[lb]{\smash{{\SetFigFont{5}{6.0}{\familydefault}{\mddefault}{\updefault}{\color[rgb]{0,0,0}$y_1$}%
}}}}
\put(5851,-3136){\makebox(0,0)[lb]{\smash{{\SetFigFont{5}{6.0}{\familydefault}{\mddefault}{\updefault}{\color[rgb]{0,0,0}$u_2$}%
}}}}
\put(4951,-2686){\makebox(0,0)[lb]{\smash{{\SetFigFont{5}{6.0}{\familydefault}{\mddefault}{\updefault}{\color[rgb]{0,0,0}$G$}%
}}}}
{\color[rgb]{0,0,0}\put(3901,-3661){\framebox(2400,2100){}}
}%
\end{picture}%

\caption{Quantum feedback network described by the linear fractional transformation  $F(G)$.}
\label{fig:nw-elim1}
\end{center}
\end{figure}

Next, consider a system $G$ of the form
\begin{equation}
G = \left(
\left(  \begin{array}{cc}
S_{11} & S_{12}
\\
S_{21}  & S_{22}
\end{array}  \right),
\left( \begin{array}{c}
L_1
\\
L_2
\end{array}  \right),
H
\right) .
\label{G-2x2}
\end{equation}
The feedback network $F(G)$ defined by Figure \ref{fig:nw-elim1}
is well-defined provided $I-S_{22}$ is invertible, in which case
the parameters for $F(G)$ are given by
  the linear fractional transformation  \cite{GJ08a}
\begin{eqnarray}
F(G) = (
S_{11}+ S_{12}(I-S_{22})^{-1} S_{21}, \
L_1+ S_{12}(I-S_{22})^{-1} L_2,
\nonumber \\
H+ \mathrm{Im} \{  L_1^\dagger S_{12}(I-S_{22})^{-1} L_2 \}
+ \mathrm{Im} \{  L_2^\dagger S_{22}(I-S_{22})^{-1} L_2 \}
) .
\label{G-fb}
\end{eqnarray}

 \begin{remark}
Underlying the series and LFT network constructs is the simple
idea of equating the  input $u_2$ with a slightly delayed version of the
output $y_2$,  and then letting the delay tend to zero. Full technical
details are given in \cite{GZ00}, \cite{GJ08}, \cite{GJ08a}.
 $\Box$
 \end{remark}

\section{Quantum Dissipative Systems}
\label{sec:qdiss}

We are now in a position to introduce a general definition of {\em
dissipation} for open quantum systems. As we have indicated above,
since dissipation concerns the loss of energy or coherence from a
system of interest, which we call the {\em plant}, to an external
environment, or the effect of an external system or environment on
the system,   we model the external influences as another open
system, which we call an {\em exosystem}. Since we wish to
consider the effect of a range of exosystems (analogous to a range
of signal generators connected to an RLC circuit), we specify a
class of allowed exosystems for the dissipation property.  The
definition, together with some examples, is presented in
subsection \ref{sec:qdiss-dfn} using the QFN framework from
section \ref{sec:open-qfn}.

\subsection{Definitions and Differential Characterization}
\label{sec:qdiss-dfn}

\begin{figure}[h]
\begin{center}

\setlength{\unitlength}{1379sp}%
\begingroup\makeatletter\ifx\SetFigFont\undefined%
\gdef\SetFigFont#1#2#3#4#5{%
  \reset@font\fontsize{#1}{#2pt}%
  \fontfamily{#3}\fontseries{#4}\fontshape{#5}%
  \selectfont}%
\fi\endgroup%
\begin{picture}(7244,5144)(1479,-6683)
\put(4126,-3136){\makebox(0,0)[lb]{\smash{{\SetFigFont{5}{6.0}{\familydefault}{\mddefault}{\updefault}{\color[rgb]{0,0,0}2}%
}}}}
\thicklines
{\color[rgb]{0,0,0}\put(7801,-2161){\vector(-1, 0){1500}}
}%
{\color[rgb]{0,0,0}\put(3001,-2161){\vector(-1, 0){1500}}
}%
{\color[rgb]{0,0,0}\put(2851,-2161){\line( 1, 0){1050}}
}%
{\color[rgb]{0,0,0}\put(7726,-2161){\line( 1, 0){975}}
}%
{\color[rgb]{0,0,0}\put(3901,-6661){\framebox(2400,2100){}}
}%
{\color[rgb]{0,0,0}\put(5101,-3661){\vector( 0, 1){  0}}
\put(5101,-3661){\vector( 0,-1){900}}
}%
{\color[rgb]{0,0,0}\put(2401,-6061){\vector( 1, 0){1500}}
}%
{\color[rgb]{0,0,0}\put(2476,-6061){\line(-1, 0){975}}
}%
{\color[rgb]{0,0,0}\put(7351,-6061){\line(-1, 0){1050}}
}%
{\color[rgb]{0,0,0}\put(7201,-6061){\vector( 1, 0){1500}}
}%
{\color[rgb]{0,0,0}\put(3901,-3061){\line(-1, 0){1500}}
\put(2401,-3061){\line( 0,-1){2100}}
\put(2401,-5161){\vector( 1, 0){1500}}
}%
{\color[rgb]{0,0,0}\put(6301,-5161){\line( 1, 0){1500}}
\put(7801,-5161){\line( 0, 1){2100}}
\put(7801,-3061){\vector(-1, 0){1500}}
}%
\put(5026,-5611){\makebox(0,0)[lb]{\smash{{\SetFigFont{5}{6.0}{\familydefault}{\mddefault}{\updefault}{\color[rgb]{0,0,0}$P$}%
}}}}
\put(5026,-2611){\makebox(0,0)[lb]{\smash{{\SetFigFont{5}{6.0}{\familydefault}{\mddefault}{\updefault}{\color[rgb]{0,0,0}$W$}%
}}}}
\put(6001,-6136){\makebox(0,0)[lb]{\smash{{\SetFigFont{5}{6.0}{\familydefault}{\mddefault}{\updefault}{\color[rgb]{0,0,0}2}%
}}}}
\put(4126,-6136){\makebox(0,0)[lb]{\smash{{\SetFigFont{5}{6.0}{\familydefault}{\mddefault}{\updefault}{\color[rgb]{0,0,0}2}%
}}}}
\put(4126,-5236){\makebox(0,0)[lb]{\smash{{\SetFigFont{5}{6.0}{\familydefault}{\mddefault}{\updefault}{\color[rgb]{0,0,0}1}%
}}}}
\put(6001,-5236){\makebox(0,0)[lb]{\smash{{\SetFigFont{5}{6.0}{\familydefault}{\mddefault}{\updefault}{\color[rgb]{0,0,0}1}%
}}}}
\put(6001,-3136){\makebox(0,0)[lb]{\smash{{\SetFigFont{5}{6.0}{\familydefault}{\mddefault}{\updefault}{\color[rgb]{0,0,0}2}%
}}}}
\put(6001,-2236){\makebox(0,0)[lb]{\smash{{\SetFigFont{5}{6.0}{\familydefault}{\mddefault}{\updefault}{\color[rgb]{0,0,0}1}%
}}}}
\put(4126,-2236){\makebox(0,0)[lb]{\smash{{\SetFigFont{5}{6.0}{\familydefault}{\mddefault}{\updefault}{\color[rgb]{0,0,0}1}%
}}}}
{\color[rgb]{0,0,0}\put(3901,-3661){\framebox(2400,2100){}}
}%
\end{picture}%

\caption{The plant-exosystem network $P \wedge W$.}
\label{fig:pw}
\end{center}
\end{figure}
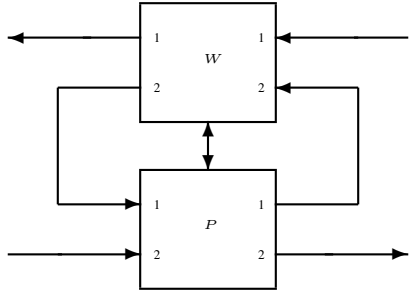

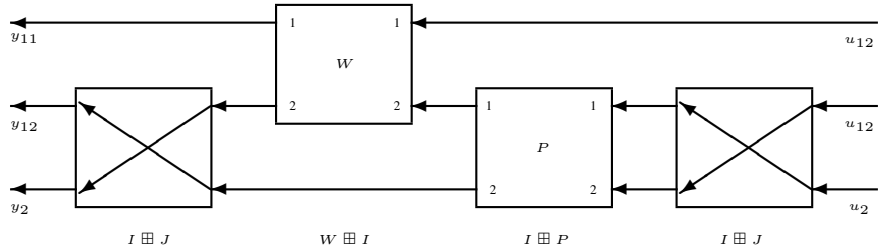
\begin{figure}[h]
\begin{center}

\setlength{\unitlength}{1379sp}%
\begingroup\makeatletter\ifx\SetFigFont\undefined%
\gdef\SetFigFont#1#2#3#4#5{%
  \reset@font\fontsize{#1}{#2pt}%
  \fontfamily{#3}\fontseries{#4}\fontshape{#5}%
  \selectfont}%
\fi\endgroup%
\begin{picture}(15644,4375)(-921,-5914)
\put(4951,-2686){\makebox(0,0)[lb]{\smash{{\SetFigFont{5}{6.0}{\familydefault}{\mddefault}{\updefault}{\color[rgb]{0,0,0}$W$}%
}}}}
\thicklines
{\color[rgb]{0,0,0}\put(7501,-5161){\framebox(2400,2100){}}
}%
{\color[rgb]{0,0,0}\put(301,-5161){\framebox(2400,2100){}}
}%
{\color[rgb]{0,0,0}\put(11101,-5161){\framebox(2400,2100){}}
}%
{\color[rgb]{0,0,0}\put(301,-3361){\vector(-1, 0){1200}}
}%
{\color[rgb]{0,0,0}\put(301,-4861){\vector(-1, 0){1200}}
}%
{\color[rgb]{0,0,0}\put(7501,-3361){\vector(-1, 0){1200}}
}%
{\color[rgb]{0,0,0}\put(11101,-3361){\vector(-1, 0){1200}}
}%
{\color[rgb]{0,0,0}\put(11101,-4861){\vector(-1, 0){1200}}
}%
{\color[rgb]{0,0,0}\put(14701,-4861){\vector(-1, 0){1200}}
}%
{\color[rgb]{0,0,0}\put(14701,-3361){\vector(-1, 0){1200}}
}%
{\color[rgb]{0,0,0}\put(3901,-1861){\vector(-1, 0){4800}}
}%
{\color[rgb]{0,0,0}\put(3901,-3361){\vector(-1, 0){1200}}
}%
{\color[rgb]{0,0,0}\put(2701,-3361){\vector(-3,-2){2353.846}}
}%
{\color[rgb]{0,0,0}\put(2701,-4861){\vector(-3, 2){2353.846}}
}%
{\color[rgb]{0,0,0}\put(13501,-4861){\vector(-3, 2){2353.846}}
}%
{\color[rgb]{0,0,0}\put(13501,-3361){\vector(-3,-2){2353.846}}
}%
{\color[rgb]{0,0,0}\put(7501,-4861){\vector(-1, 0){4800}}
}%
{\color[rgb]{0,0,0}\put(14701,-1861){\vector(-1, 0){8400}}
}%
\put(6001,-3436){\makebox(0,0)[lb]{\smash{{\SetFigFont{5}{6.0}{\familydefault}{\mddefault}{\updefault}{\color[rgb]{0,0,0}2}%
}}}}
\put(6001,-1936){\makebox(0,0)[lb]{\smash{{\SetFigFont{5}{6.0}{\familydefault}{\mddefault}{\updefault}{\color[rgb]{0,0,0}1}%
}}}}
\put(4126,-1936){\makebox(0,0)[lb]{\smash{{\SetFigFont{5}{6.0}{\familydefault}{\mddefault}{\updefault}{\color[rgb]{0,0,0}1}%
}}}}
\put(4126,-3436){\makebox(0,0)[lb]{\smash{{\SetFigFont{5}{6.0}{\familydefault}{\mddefault}{\updefault}{\color[rgb]{0,0,0}2}%
}}}}
\put(7651,-3436){\makebox(0,0)[lb]{\smash{{\SetFigFont{5}{6.0}{\familydefault}{\mddefault}{\updefault}{\color[rgb]{0,0,0}1}%
}}}}
\put(7726,-4936){\makebox(0,0)[lb]{\smash{{\SetFigFont{5}{6.0}{\familydefault}{\mddefault}{\updefault}{\color[rgb]{0,0,0}2}%
}}}}
\put(8551,-4186){\makebox(0,0)[lb]{\smash{{\SetFigFont{5}{6.0}{\familydefault}{\mddefault}{\updefault}{\color[rgb]{0,0,0}$P$}%
}}}}
\put(9526,-3436){\makebox(0,0)[lb]{\smash{{\SetFigFont{5}{6.0}{\familydefault}{\mddefault}{\updefault}{\color[rgb]{0,0,0}1}%
}}}}
\put(9526,-4936){\makebox(0,0)[lb]{\smash{{\SetFigFont{5}{6.0}{\familydefault}{\mddefault}{\updefault}{\color[rgb]{0,0,0}2}%
}}}}
\put(1201,-5836){\makebox(0,0)[lb]{\smash{{\SetFigFont{5}{6.0}{\familydefault}{\mddefault}{\updefault}{\color[rgb]{0,0,0}$I \boxplus J$}%
}}}}
\put(4651,-5836){\makebox(0,0)[lb]{\smash{{\SetFigFont{5}{6.0}{\familydefault}{\mddefault}{\updefault}{\color[rgb]{0,0,0}$W \boxplus I$}%
}}}}
\put(8326,-5836){\makebox(0,0)[lb]{\smash{{\SetFigFont{5}{6.0}{\familydefault}{\mddefault}{\updefault}{\color[rgb]{0,0,0}$I \boxplus P$}%
}}}}
\put(11851,-5836){\makebox(0,0)[lb]{\smash{{\SetFigFont{5}{6.0}{\familydefault}{\mddefault}{\updefault}{\color[rgb]{0,0,0}$I \boxplus J$}%
}}}}
\put(-899,-2161){\makebox(0,0)[lb]{\smash{{\SetFigFont{5}{6.0}{\familydefault}{\mddefault}{\updefault}{\color[rgb]{0,0,0}$y_{11}$}%
}}}}
\put(-899,-3736){\makebox(0,0)[lb]{\smash{{\SetFigFont{5}{6.0}{\familydefault}{\mddefault}{\updefault}{\color[rgb]{0,0,0}$y_{12}$}%
}}}}
\put(-899,-5236){\makebox(0,0)[lb]{\smash{{\SetFigFont{5}{6.0}{\familydefault}{\mddefault}{\updefault}{\color[rgb]{0,0,0}$y_{2}$}%
}}}}
\put(14176,-5236){\makebox(0,0)[lb]{\smash{{\SetFigFont{5}{6.0}{\familydefault}{\mddefault}{\updefault}{\color[rgb]{0,0,0}$u_{2}$}%
}}}}
\put(14101,-3736){\makebox(0,0)[lb]{\smash{{\SetFigFont{5}{6.0}{\familydefault}{\mddefault}{\updefault}{\color[rgb]{0,0,0}$u_{12}$}%
}}}}
\put(14101,-2236){\makebox(0,0)[lb]{\smash{{\SetFigFont{5}{6.0}{\familydefault}{\mddefault}{\updefault}{\color[rgb]{0,0,0}$u_{12}$}%
}}}}
{\color[rgb]{0,0,0}\put(3901,-3661){\framebox(2400,2100){}}
}%
\end{picture}%

\caption{The  system $G$ used to represent  the network $P \wedge W$ of Figure \ref{fig:pw}.}
\label{fig:pw-g}
\end{center}
\end{figure}

A {\em plant} $ {P}$ is an open system of the type defined in subsection \ref{sec:open-defs}:
\begin{equation}
 {P} = ( {S},  {L}, H).
\label{plant-dfn}
\end{equation}
It is the system of main interest, and is regarded as being part
of a possibly larger network of systems. The space of physical
variables for  $ {P}$ is  denoted $\mathscr{A}_{ {P}}$. Since our
interest  is in modeling and analyzing the effect of the external
influences  on the plant, we assume that  certain of its input and
outputs are available   for connection to an  {\em exosystem} $
{W}$, an open system
\begin{equation}
 {W} =( {R},  {w}, D) .
\label{exo-dfn}
\end{equation}
It is also assumed that certain plant variables are available for
direct interconnection with an exosystem. Thus the plant
definition and interconnection specification determine how it can
interface with exosystems $ {W}$, as discussed in subsection
\ref{sec:open-qfn}. The  interconnections determine a network $
{P} \wedge  {W}$, as shown in Figure \ref{fig:pw} (a star product
architecture, \cite{GL95}, \cite{ZDG96}, \cite{GJ08a}). In terms
of the network constructs of section \ref{sec:open-qfn}, we have
explicitly
\begin{equation}
P \wedge W = F(   G)  \boxplus H_{PW}  ,
 \label{P-wedge-W}
\end{equation}
using the LFT (Figure \ref{fig:nw-elim1}), where 
\begin{equation}
 G= (I \boxplus J) \triangleleft (W \boxplus I) \triangleleft ( I \boxplus P) \triangleleft (I \boxplus J),
  \label{P-wedge-W-G}
\end{equation}
as shown in Figure \ref{fig:pw-g}, and
$H_{PW}$ is a direct interaction Hamiltonian.

We will allow exosystems $ {W}$ to vary in a class $\mathscr{W}$
of such exosystems. The operators constituting the system
parameters $ {W}$ belong to an algebra $\mathscr{A}_{ {P}} \otimes
\mathscr{A}_{ex}$.
 The input to the system $ {P} \wedge  {W}$ is assumed to be a canonical vacuum field.

Let $r=r_{ {P}}(  {W} ) \in \mathscr{A}_{ {P}} \otimes
\mathscr{A}_{ex}$ be a self-adjoint symmetrically ordered function
of the exosystem parameters, depending on the plant parameters
(e.g. a quadratic form), called a {\em supply rate}.  In general,
a supply rate may contain constant terms (see Example
\ref{eg:cavity-1}).

We say that the plant  $ {P}$ is {\em dissipative} with supply rate $r$
with respect to a class $\mathscr{W}$ of exosystems
if there exists a non-negative plant  observable $V\in \mathscr{A}_{ {P}}$   such that
\begin{equation}
\mathbb{E}_0 \left[ V(t) -  V   - \int_0^t   r( {W})(s)   ds \right]  \leq 0
\label{quant-diss-1}
\end{equation}
for all  exosystems $ {W} \in \mathscr{W}$  and  all  $t \geq 0$.
Here, the dynamics is that of the  network $ {P} \wedge  {W}$,
$\mathbb{E}_0$ is the initial vacuum expectation onto
$\mathscr{A}_{ {P}} \otimes \mathscr{A}_{ex}$ (which contains
$\mathscr{A}_{ {P}}$), and $\leq$ denotes operator ordering, see
Appendix \ref{sec:app-order}.

We shall refer to (\ref{quant-diss-1}) as the {\em dissipation
inequality}. It is a natural generalization of Willems' definition
\cite{WIL72} to open quantum systems, building on \cite{DJP00},
\cite{JNP08}. Following the terminology from \cite{WIL72}, when
equality holds in (\ref{quant-diss-1})    for all exosystems  $
{W} \in \mathscr{W}$, we say that the plant $ {P}$ is {\em
lossless}. This terminology refers to the fact that in such cases
all energy flows and storages have been accounted for; the total
system (plant, exosystem, fields) does not loose energy.
 Important special cases and applications of the dissipation inequality will be discussed below.

We next present the infinitesimal version of the dissipation
inequality (\ref{quant-diss-1}) in the following theorem.
%Recall that in quantum mechanics an {\em observable} is represented as a self-adjoint operator.

\begin{theorem}  \label{thm:qdiss-diff}
Let $P$ be a plant, and $P \wedge W$ be a given network structure
for a class $\mathscr{W}$ of exosystems. Then the plant  $ {P}$ is
{\em dissipative} with supply rate $r$ with respect to
$\mathscr{W}$ if and only if  there exists a non-negative system
observable $V\in \mathscr{A}_{ {P}}$   such that
\begin{eqnarray}
 \mathcal{G}_{  {P} \wedge  {W} }(V)
 -  r( {W})   \leq 0
\label{quant-diss-2}
\end{eqnarray}
for all exosystem parameters $ {W}  \in \mathscr{W}$.
\end{theorem}

\begin{proof}
Assume $ {P}$ is dissipative as stated,
and select any exosystem $ {W}\in \mathscr{W}$ and consider the plant-exosystem network $ {P} \wedge  {W}$.
Now apply (\ref{vac-exp}) to the network to obtain
\begin{eqnarray}
\frac{d}{dt} \mathbb{E}_0 [ V(t) ] \vert_{t=0} &=&
\mathcal{G}_{ {P} \wedge  {W}}(V) .
\label{P<Lddt}
\end{eqnarray}
Now combine this with (\ref{quant-diss-1}) to obtain
(\ref{quant-diss-2}). The converse follows by reversing this
argument and integration.
\end{proof}

Note that the dissipation inequality (\ref{quant-diss-2}) involves
only plant and exosystem operators, since the LHS depends on
operators  in  the  algebra $\mathscr{A}_{ {P}} \otimes
\mathscr{A}_{ex}$  (the noise has been averaged out by
$\mathbb{E}_0$).

\begin{remark}  \label{rmk:time-dep-1}
(Non-autonomous case.) For non-autonomous situations, the plant
and exosystem operators may depend on time (they can even be
random provided they are adapted), and the dissipation
inequalities take the forms
\begin{equation}
\mathbb{E}_s \left[ V(t) -  V(s)   - \int_s^t   r(  {W})(r)   dr \right]   \leq 0
\label{quant-diss-1-t}
\end{equation}
for all $t \geq s$ and all exosystems  $ {W} \in \mathscr{W}$, and
\begin{eqnarray}
    \mathcal{G}_{  {P} \wedge  {W}  }(V)(t)
 -  r( {W})(t)  \leq 0
\label{quant-diss-2-t}
\end{eqnarray}
for all $t \geq 0$ and all exosystem parameters $ {W} \in \mathscr{W}$.
$\Box$
\end{remark}

In general we can expect physical systems to be dissipative in
some sense, and indeed we characterize this precisely in the next
theorem.  In what follows we denote by $\mathscr{W}_u$ the class
of exosystems without any particular constraints
(unrestricted)---the only requirement is that members of
$\mathscr{W}_u$ are open systems that can be connected to the
plant. The following simple (but important!) result follows from a
calculation similar to that used to prove Theorem
\ref{thm:qdiss-diff}.

\begin{theorem}  \label{thm:qdiss-general}
Let $ {P}=( {S},  {L}, H)$ be a plant, and $P \wedge W$ be a given
network structure for the class $\mathscr{W}_u$ of (unconstrained)
exosystems. Assume there exists an observable $V_0 \geq 0$
commuting with $H$. Then $ {P}$ is lossless (hence dissipative)
with respect to $\mathscr{W}_u$ for a supply rate $r=r_0$  given
by
\begin{equation}
r_0( {W}) = \mathcal{G}_{ {P}  \wedge  {W}}(V_0) ,
\label{qdiss-general-r-1}
\end{equation}
with storage function $V=V_0$.
\end{theorem}

Theorems \ref{thm:qdiss-diff} and \ref{thm:qdiss-general} are
quite general, and contain   many special cases of interest, e.g.
versions of  the Bounded and Positive Real Lemmas, as well as
fundamental energy balance properties of physical systems. See
Theorems \ref{thm:prl} and \ref{thm:brl} below.

It is instructive to examine the form of that
\lq\lq{natural}\rq\rq \ supply rate $r_0$ (defined by
(\ref{qdiss-general-r-1})) for the specific interconnection
between the plant $P=(I,L,H)$ and exosystems given by the series
connection
\begin{equation}
 {P} \wedge  {W} =  {P} \triangleleft  {W} ,
\label{NPW-series}
\end{equation}
and where the Hamiltonian $D$ of the exosystem has the form
\begin{equation}
D=-i (  {K}^\dagger  {v} - {v}^\dagger  {K} ) ,
\label{D-std}
\end{equation}
where $ {K} \in \mathscr{A}_{ {P}}$ is fixed and $ {v}$ commutes with $\mathscr{A}_{ {P}}$.
Then we have
\begin{eqnarray}
r_0( {W}) &=& r_0( {w}, {v})= \mathcal{L}_{ {w}} (V_0) +
\mathcal{L}_{ {L}} (V_0) + {w}^\dagger  [V_0,  {L}] + [
{L}^\dagger, V_0]   {w}  -i [ V_0, D]
\nonumber \\
&=& \mathcal{L}_{ {w}} (V_0) + \mathcal{L}_{ {L}} (V_0) + \left(
\begin{array}{cc}  {w}^\dagger &  {v}^\dagger \end{array} \right)
[V_0, \left( \begin{array}{c}   {L}  \\  {K} \end{array} \right)]
+ [\left( \begin{array}{cc}  {L}^\dagger &  {K}^\dagger
\end{array} \right) , V_0]  \left( \begin{array}{c}  {w} \\  {v}
\end{array} \right)
\nonumber \\
&=&  \mathcal{L}_{ {w}} (V_0) + \mathcal{L}_{ {L}} (V_0) +
\left( \begin{array}{cc}  {w}^\dagger &  {v}^\dagger \end{array} \right)
  {Z}
+  {Z}^\dagger  \left( \begin{array}{c}  {w} \\  {v} \end{array} \right)  ,
\label{qdiss-general-r-2}
\end{eqnarray}
where
\begin{equation}
 {Z}=[V_0, \left( \begin{array}{c}   {L}  \\  {K} \end{array} \right)] .
\label{Z-def-general}
\end{equation}
This means that the plant $P=(I,L,H)$ satisfies the lossless energy rate relation
\begin{equation}
\mathcal{G}_{P \triangleleft W}(V_0) =  \mathcal{L}_{ {w}} (V_0) + \mathcal{L}_{ {L}} (V_0) +
\left( \begin{array}{cc}  {w}^\dagger &  {v}^\dagger \end{array} \right)
  {Z}
+  {Z}^\dagger  \left( \begin{array}{c}  {w} \\  {v} \end{array} \right) .
\label{P-lossless}
\end{equation}
This relation may be regarded as a generalization of the
well-known passivity relation for classical Hamiltonian systems
(see, e.g. \cite[eq. (2.39) and Chapter 4]{VDS96}).\footnote{For a
classical Hamiltonian system with Hamiltonian $H=H_0+Gu$, we have
$dH_0/dt = \{ H_0, H \} = z u$, where $u$ is an input   and $z
=\{H_0, G\}$ is an output (here the Poisson bracket is defined by
$\{ f, g \} = \nabla^T f \Sigma \nabla g$, where $\Sigma=\left(
\begin{array}{cc} 0 & 1 \\ -1 & 0 \end{array} \right)$).} If we
think of $ {Z}$ as an \lq\lq{output}\rq\rq \ quantity (not
necessarily part of an output field channel!), then the last two
terms correspond to a product of input and output terms, i.e. an
energy flow into the plant from the exosystem---these terms
capture the {\em passivity} of the plant, as we now explain. The
term $\mathcal{L}_{ {L}} (V_0)$ is known in quantum physics as a
\lq\lq{dissipation}\rq\rq \ or \lq\lq{damping}\rq\rq\ \ term,
which in the model arises from the quantum noise in the field;
when negative, this term represents energy lost from the plant to
the field (heat bath). The first term, $\mathcal{L}_{ {w}} (V_0)$,
arises from a \lq\lq{$dt$}\rq\rq \ or \lq\lq{signal}\rq\rq \  term
in the field inputs, and is in general zero when $ {w}$ commutes
with plant operators; it is an exosystem partner to the field
dissipation term. Indeed, if we integrate (\ref{P-lossless}) and
take expectations with respect to a joint state consisting of a
plant state vector $\psi_0$ for which $\langle \psi_0, V_0 \psi_0
\rangle =0$ and the vacuum state for the network input fields, we
find that
\begin{equation}
\int_0^t \langle    \mathcal{L}_{ {w}} (V_0)(s) + \mathcal{L}_{ {L}} (V_0)(s) +
\left( \begin{array}{cc}  {w}^\dagger(s) &  {v}^\dagger(s) \end{array} \right)
  {Z}(s)
+  {Z}^\dagger (s)  \left( \begin{array}{c}  {w}(s) \\  {v} (s)
\end{array} \right) \rangle ds \geq 0, \label{P-passive}
\end{equation}
which generalizes well known passivity inequalities. Passivity and
gain will be discussed in more detail in section
\ref{sec:qdiss-pg},  which also contains illustrative examples.

\subsection{Stability}
\label{sec:qdiss-stability}

In the case of classical deterministic systems, dissipativity
plays an important role in stability analysis, \cite{WIL72},
\cite{HM76}. For instance, strictly passive systems  are
asymptotically stable. The purpose of this section is to make this
connection in our current quantum context using a simple
exponential stability criterion, given in the following lemma.

\begin{lemma} \label{lemma:stability}
 Let $ {P}$ be a plant for which there exists  a non-negative observable $V$  satisfying the differential inequality
\begin{equation}
\mathcal{G}_{ {P}}( V ) + c V \leq \lambda,
\label{stab-1-1}
\end{equation}
where $c > 0$ and $\lambda$ are real numbers. Then for any plant state   we have
\begin{equation}
  \langle  V(t)  \rangle \leq e^{-ct}   \langle V \rangle + \lambda/c .
\label{stab-1-2}
\end{equation}
Moreover, if $\lambda=0$ then $\lim_{t\to\infty} \langle  V(t) \rangle =0$.
\end{lemma}

\begin{proof}
Integrating the plant dynamics we find that
\begin{equation}
 \mathbb{E}_t \left[ V(t+h) - V(t) - c\int_t^{t+h} V(r) dr \right] \leq \lambda h , \ \ ( h \geq 0). 
\label{stab-2-1}
\end{equation}
Now from \cite[sec. 26]{KRP92}, we have $\mathbb{E}_0 \mathbb{E}_t =\mathbb{E}_0$ for $t\geq 0$, hence it follows that the expectation $\langle V(t) \rangle = \langle \psi,  \mathbb{E}_0[V(t)] \psi \rangle$ ($\psi$ is the plant state) satisfies
\begin{equation}
 \langle V(t+h) \rangle - \langle V(t) \rangle  - c\int_t^{t+h} \langle V(r) \rangle dr   \leq \lambda h , \ \  (h \geq 0) .
\label{stab-2-2}
\end{equation}
From this we deduce
\begin{equation}
 \frac{d}{dt} \langle V(t) \rangle \leq -c \langle V(t) \rangle +\lambda.
\label{stab-3}
\end{equation}
The assertions of the lemma now follow.
\end{proof}

We now state a stability result for strictly dissipative systems.

\begin{theorem}   \label{thm:strict-passive-stability}
Let $ {P}$ be dissipative with storage function $V$ and supply
rate $r( {W})$. Assume that the supply rate satisfies
\begin{equation}
r( I ) \leq -c V
\label{strict-passive-supply}
\end{equation}
for some real number $c>0$. Then for the unperturbed plant $ {P}=
{P} \triangleleft I $ we have $\lim_{t\to\infty} \langle  V(t)
\rangle =0$ for any plant state.
\end{theorem}

This result follows, of course, from Lemma \ref{lemma:stability}
and Theorem \ref{thm:qdiss-diff}. It is possible to consider more
general stability results and this will be taken up in future
work.

\subsection{Passivity and Gain}
\label{sec:qdiss-pg}

In this subsection we consider passivity and gain properties for
the series plant-exosystem interconnection, (\ref{NPW-series}). In
particular, we present generalizations of the positive and bounded
real lemmas (Theorems \ref{thm:prl} and \ref{thm:brl} below). To
help make the ideas clearer, we will assume that the exosystems
commute with the plant. More general results can be formulated
using the framework developed in this paper.

Consider  a plant $ {P}=( {I},  {L}, H)$, and let ${N},  {Z} \in
\mathscr{A}_{ {P}}$. Define a class of exosystems
$$
\mathscr{W}_1 = \left\{  {W} =  ( {R},  {w}, -i(  {K}^\dagger  {v} - {v}^\dagger  {K} )) \ : \  {K} \in \mathscr{A}_{ {P}} \ \mathrm{is \ fixed\ and} \   {w},  {v} \ \mathrm{commute \ with} \ \mathscr{A}_{ {P}} \right\} .
$$
We say that $ {P}$ is {\em passive} (with respect to $\mathscr{W}_1$)  if
 it is dissipative with respect to the supply rate
\begin{equation}
r( {W}) = r( {w}, {v})=   -  {N}^\dagger  {N}   +
\left( \begin{array}{cc}  {w}^\dagger &  {v}^\dagger \end{array} \right)
  {Z}
+  {Z}^\dagger  \left( \begin{array}{c}  {w} \\  {v} \end{array} \right)
  +  \lambda
\label{r-passive}
\end{equation}
for some non-negative real number $\lambda$.
 Explicitly, this means that $ {P}$ is passive if
 there exists a non-negative system observable $V\in \mathscr{A}_{ {P}}$ and a non-negative real number $\lambda$ such that
\begin{equation}
\mathbb{E}_0 \left[ V(t) -  V   - \int_0^t  (\left( \begin{array}{c}    {w}^\dagger(s)   \\  {v}^\dagger(s) \end{array} \right)  {Z}(s)    +    {Z}^\dagger(s)   \left( \begin{array}{cc}  {w}(s) &  {v}(s) \end{array} \right)   -
  {N}^\dagger(s)  {N} (s)  )  ds \right] - \lambda t \leq 0
\label{quant-diss-1-prl}
\end{equation}
for all  exosystems $ {W} \in \mathscr{W}_1$  and  all  $t \geq 0$.

\begin{theorem}   \label{thm:prl}
(Positive Real Lemma) A plant $ {P}=( {I},  {L}, H)$ is passive
(with respect to $\mathscr{W}_1$)  if and only if there exists a
non-negative system observable $V\in \mathscr{A}_{ {P}}$ and a
non-negative real number $\lambda$ such that
\begin{eqnarray}
\mathcal{L}_{ {L}}(V)
   - i[V,H  ]   + {N}^\dagger  {N}  - \lambda  \leq 0 ,
\label{prl-1} \\
 {Z}=[V, \left( \begin{array}{c}  {L} \\  {K} \end{array} \right)  ] .
\label{prl-2}
\end{eqnarray}
\end{theorem}

\begin{proof}
We use some results from the Appendix in what follows. If $ {P}$
is passive, there exists a non-negative system observable $V\in
\mathscr{A}_{ {P}}$ and a non-negative real number $\lambda$ such
that
\begin{eqnarray}
 \mathcal{L}_{ {L}}(V)    -i[V, H ]
%\nonumber \\
 + \left( \begin{array}{cc} {w}^\dagger \\
 {v}^\dagger \end{array} \right) ( [V,  {L}]- {Z}) + ([ {L}^\dagger, V]- {Z}^\dagger) \left( \begin{array}{cc}   {w} &  {v} \end{array} \right)
    +
  {N}^\dagger  {N}    -   \lambda \leq 0
\label{prl-pf-1}
\end{eqnarray}
for all exosystem parameters ${W}=( {R},  {w}, -i( {v}^\dagger
{K} -  {K}^\dagger  {v} )) \in \mathscr{W}_1$. Setting $ {w}=0$
and $ {v}=0$ we obtain (\ref{prl-1}). Now let $ {w}$ vary freely
but commuting with plant operators to deduce  (\ref{prl-2}).
\end{proof}

The Bounded Real Lemma is used to determine the $L^2$ gain of a
system, and in conjunction with the Small Gain Theorem, can be
used for robust stability analysis and design, \cite{GL95},
\cite{ZDG96}, \cite{DJ06}. Again consider a plant  $ {P}=( {1},
{L}, H)$,  let $ {N},  {Z} \in \mathscr{A}_{ {P}}$, and let $g >
0$ be a real number (gain parameter). In order to simplify the
exposition, we restrict our attention to the class
$$
\mathscr{W}_2 = \left\{  {W} =  ( {R},  {w},  0) \ : \ {w} \
\mathrm{commutes \ with} \ \mathscr{A}_{ {P}} \right\} .
$$
We say that $ {P}$ has {\em  gain   $g$} (with respect to
$\mathscr{W}_2$)  if it is dissipative with respect to the supply
rate
\begin{equation}
r(  {W}) =   g^2    {w}^\dagger  {w}
  -
 ( {N}+ {Z} {w})^\dagger ( {N} + {Z} {w}) + \lambda ,
 \label{quant-diss-fg-r}
 \end{equation}
for a  real number  $\lambda \geq 0$, and class $\mathscr{W}_2$;
i.e. there exists a non-negative system observable $V\in
\mathscr{A}_{ {P}}$ and a non-negative real number $\lambda$ such
that
\begin{equation}
\mathbb{E}_0 \left[ V(t) -  V   - \int_0^t  ( g^2  {w}^\dagger(s)  {w} (s)
  -
 ( {N}(s)+ {Z}(s) {w}(s))^\dagger ( {N} (s)+ {Z}(s) {w}(s))
)  ds \right] - \lambda t \leq 0
\label{quant-diss-1-prl-2}
\end{equation}
for all exosystems $ {W} \in \mathscr{W}_2$  and  all  $t \geq 0$.

\begin{theorem}   \label{thm:brl}
(Bounded Real Lemma) A plant $ {P}=( {I},  {L}, H)$ has  gain $g$
(with respect to $\mathscr{W}_2$) if and only if there exists a
non-negative system observable $V\in \mathscr{A}_{ {P}}$ and a
non-negative real number $\lambda$ such that
\begin{eqnarray}
 {\Gamma} =  g^2 - {Z}^\dagger  {Z}  \geq   0 \
 \label{brl-Gamma}
 \end{eqnarray}
 and
 \begin{eqnarray}
   \mathcal{L}_{ {L}}(V)    -i[V, H ]
%\nonumber \\
 - {w}^\dagger  {\Gamma}  {w}
  +  {w}^\dagger ([V, {L}] +  {Z}^\dagger  {N} )
 + ([V, {L}] +  {Z}^\dagger  {N} )^\dagger  {w}
 + {N}^\dagger  {N}
    -   \lambda \leq 0
\label{brl-0}
\end{eqnarray}
 for all exosystem parameters $ {w}$. If $ {\Gamma}^{-1}$ exists, then $ {P}=( {1}, {L}, H)$ has  gain $g$  (with respect to $\mathscr{W}_2$) if
  \begin{eqnarray}
 \mathcal{L}_{ {L}}(V) -i [V,H] + {N}^\dagger  {N}
 +([ {L}^\dagger, V] +  {N}^\dagger  {Z})  {\Gamma}^{-1} (  [V,  {L}] +  {Z}^\dagger  {N})
 - \lambda \leq 0 .
\label{brl-2-c}
\end{eqnarray}
 \end{theorem}

\begin{proof}
If $ {P}$ has gain $g$, then there exists a non-negative system
observable $V\in \mathscr{A}_{ {P}}$ and  a real number $
\lambda\geq 0 $ such that
\begin{eqnarray}
   \mathcal{L}_{ {L}}(V)    -i[V, H ]
%\nonumber \\
 - {w}^\dagger  {\Gamma}   {w}
  +  {w}^\dagger ([V, {L}] +  {Z}^\dagger  {N} )
 + ([V, {L}] +  {Z}^\dagger  {N} )^\dagger  {w}
 + {N}^\dagger  {N}
    -   \lambda \leq 0
\label{brl-pf-1}
\end{eqnarray}
for all exosystem parameters $ {w}$. From  the Appendix   we see that $ {\Gamma}   \geq 0$, hence (\ref{brl-Gamma}). In equality (\ref{brl-0}) also follows.

Next, if $ {\Gamma}^{-1}$  exists,  define
\begin{equation}
 {w}^\star =  {\Gamma}^{-1}( [V,  {L}] +  {Z}^\dagger  {N}) .
\label{w-star}
\end{equation}
Then by completion of squares
\begin{eqnarray*}
-  {w}^\dagger  {\Gamma}  {w} +  {w}^\dagger (  [V,  {L}] +  {Z}^\dagger  {N}) + ([ {L}^\dagger, V] +  {N}^\dagger  {Z}) {w}
\\
= ([ {L}^\dagger, V]+  {N}^\dagger  {Z})  {\Gamma}^{-1} (  [V,  {L}] +  {Z}^\dagger  {N}) - ( {w}- {w}^\star)^\dagger  {\Gamma}( {w}- {w}^\star)
\\
\leq
([ {L}^\dagger, V] +  {N}^\dagger  {Z})  {\Gamma}^{-1} (  [V,  {L}] +  {Z}^\dagger  {N}).
\end{eqnarray*}
This inequality and (\ref{brl-2-c}) imply (\ref{brl-0}) as required.
\end{proof}

\begin{remark} \label{rmk:brl-c}
  Note that  the \lq\lq{optimal}\rq\rq \ exosystem parameter $ {w}^\star$ defined by (\ref{w-star})
  belongs to $\mathscr{A}_{ {P}}$, and so does not in general commute with $\mathscr{A}_{ {P}}$, and consequently lies outside the class of exosystems under consideration;
i.e., $ {w}^\star  \not\in \mathscr{W}_2$ in general.
$\Box$
\end{remark}

We conclude this section with several examples.
The first example is that of a two-level system, Appendix \ref{sec:app-qbit}.

\begin{example}   \label{eg:tla-1}
(Two-level atom)
In this example we
consider an open two level atom
$$
 {P} = (1, \, \sqrt{\gamma} \, \sigma_-,  \ \frac{1}{2}\omega \sigma_z) ,
$$
where $\sigma_x$, $\sigma_y$, $\sigma_z$ denote the Pauli matrices
(Appendix \ref{sec:app-qbit}) and
$\sigma_{\pm}=\frac{1}{2}(\sigma_x \pm i\sigma_y)$.

Consider the storage function $V_0=\sigma_1 = \frac{1}{2}(I + \sigma_z) \geq 0$ (here $I$ is the $2\times 2$ identity matrix).
Then by  (\ref{qdiss-general-r-2}) we have
\begin{eqnarray}
r_0(W) &=&  -\gamma \sigma_1 - \sqrt{\gamma} (w^\ast \sigma_- + \sigma_+ w) + \mathcal{L}_w(\sigma_1) -i[\sigma_1, D]
\nonumber \\
&=& -( \sqrt{\gamma} \, \sigma_- + w)^\ast (\sqrt{\gamma} \, \sigma_- + w) + w^\ast w + \mathcal{L}_w(\sigma_1) -i[\sigma_1, D]  .
 \label{eq:tla-1}
\end{eqnarray}
Therefore this system is lossless (passive), and has gain 1  with respect to the output quantity $ \sqrt{\gamma} \, \sigma_- + w$ and commuting inputs $w$.  When $D=0$ and $ {W}=(1,0,0)$, then by Theorem \ref{thm:strict-passive-stability} the expected value of $V_0(t)=\sigma_1(t)$ tends to zero (asymptotically stable).
$\Box$
\end{example}

The next example looks at open quantum oscillators, Appendix \ref{sec:app-qho}.

\begin{example}   \label{eg:cavity-1}
(Open Oscillator)
 Consider the plant
$$
 {P} = (1, \, \alpha a + \beta a^\ast \, , \omega a^\ast a) ,
$$
where $a$ is the annihilation operator (satisfying the commutation relations $[a,a^\ast]=1$, recall Appendix \ref{sec:app-qho}),  $\alpha$ and $\beta$ are complex numbers, and $\omega$ is a frequency parameter.  The case $\alpha = \sqrt{\gamma}$, $\beta=0$ corresponds to a damped cavity with coupling $L=\sqrt{\gamma}\, a$, while the undamped oscillator model for an atom using in \cite{DJ99} has coupling $L=\sqrt{\kappa}\,(a+a^\ast)$ for which $\alpha = \beta= \sqrt{\kappa}$. The coupling $L=a^\ast$ arises in amplifier models, \cite{GZ00}.

With $V_0=a^\ast a$ and $W=(1,w,0)$, from (\ref{qdiss-general-r-2}) we have
\begin{equation}
r_0(W) = \mathscr{L}_w(V_0) + (\vert \beta \vert^2 - \vert \alpha \vert^2)V_0 + w^\ast Z + Z^\ast w + \vert \beta \vert^2 ,
\label{r0-oscillator}
\end{equation}
where $Z=-\alpha a + \beta a^\ast$. From this it can be seen that $P$ is passive whenever $\vert \beta \vert^2 - \vert \alpha \vert^2 \leq 0$.
Furthermore, if $\vert \beta \vert^2 - \vert \alpha \vert^2<0$ (strict passivity) and $W=(1,0,0)$  it follows from Lemma \ref{lemma:stability}  that the plant is stable, i.e. the expected value of $V_0$ remains bounded. If in addition $\beta=0$ then the expected value of $V_0$ tends to zero.

In the strictly passive case we may complete the squares to deduce that the plant has finite gain. For instance, if $\alpha = \sqrt{\gamma}$, $\beta=0$ (damped cavity) we have
\begin{eqnarray}
 r_0( {W}) = -(\sqrt{\gamma}\, a +w)^\ast( \sqrt{\gamma}\, a +w) + w^\ast w +\mathcal{L}_w(V_0)
\label{eg:cavity-1-4}
\end{eqnarray}
and hence the system has gain 1 relative to the output quantity $\sqrt{\gamma} \, a +w$ and commuting inputs $w$.

If $\vert \beta \vert^2 - \vert \alpha \vert^2 > 0$, the plant is not passive and not stable. The case $\vert \beta \vert^2 - \vert \alpha \vert^2 =0$ is marginally stable.
$\Box$
\end{example}

The final example of this section concerns classical deterministic
systems, of which the RLC circuit discussed earlier is a special
case (see section \ref{sec:open-prelim}).

\begin{example}   \label{eg:RLC-1}
(classical deterministic systems)
Consider the classical deterministic open system (see \cite{WIL72}, \cite{VDS96})
\begin{eqnarray}
\dot x &=& f(x) + g(x) w
\nonumber \\
z &=& h(x) ,
\label{classical-det-1}
\end{eqnarray}
where $x\in \mathbb{R}^n$, $w \in L^2_{loc}([0,\infty), W)$, $W
\subset \mathbb{R}$, $f$ and $g$ are smooth vector fields, and $h$
is a smooth real-valued function. This system is dissipative with
respect to a supply rate $r(w,z)$ if there exists a non-negative
function (the storage function) such that
\begin{equation}
V(x(t)) - V(x(0)) - \int_0^t r(w(s),z(s)) ds \leq 0
\label{classical-det-2}
\end{equation}
for all $w \in L^2[0,t]$ and all $t \geq 0$. The infinitesimal
version of this dissipation inequality is
\begin{equation}
\sup_{w \in W} \{ [f(x)+g(x)w]^T\nabla V(x) -   r(w,h(x))  \} \leq 0.
\label{classical-det-3}
\end{equation}
This is a special case of the quantum formulation given above. To
see this, we represent the classical open system as a commutative
subsystem of a quantum open system $ {P}$ with an associated
(non-autonomous) exosystem class $\mathscr{W}$. As in
\cite[Appendix B]{GJ08},
 we take the underlying Hilbert space of the classical
system to be $\mathsf{H}=L^{2}\left( \mathbb{R}^{n}\right) $ with $q^{j}$, $%
p_{j}$ being the usual canonical position and momentum observables: $%
q^{j}\psi \left(x \right) =x^{j}\psi \left( x \right) $ and $%
p_{j}\psi \left( x \right) =-i\partial _{j}\psi \left(
\vec{x}\right)$. We write $q=(q^1,\ldots,q^n)^T$,
$p=(p_1,\ldots,p_n)^T$, and $\nabla = (\partial
_{1},\ldots,\partial _{n})^T$, and define Hamiltonians
\begin{equation}
H_f=\frac{1}{2}\left(  f^T p+p^T f\right), \ \ H_g=\frac{1}{2}\left(  g^T p+p^T g\right).
\label{classical-det-4}
\end{equation}
Now if $\varphi=\varphi(q)$ is a function only of the position variables, then
\begin{equation}
 -i [ \varphi, H_f+H_g w] =  [f+gw] ^T\nabla \varphi.
\label{classical-det-5}
\end{equation}
Hence the classical open system  can be described by $ {P} \boxplus   {W}$, where
\begin{eqnarray}
  {P} &=&  (\_,\_, H_f) ,
 \nonumber \\
  {W} & \in & \mathscr{W} = \left\{      {W} =(\_,\_,H_g w) \ :  \ w \in   L^2_{loc}([0,\infty), W)
 \right\} ,
\label{classical-det-6}
\end{eqnarray}
and supply rate
\begin{equation}
 r( {W}) = r(h,w) .
\label{classical-det-7}
\end{equation}
The classical deterministic dissipation inequality
(\ref{classical-det-3}) now follows from (\ref{quant-diss-2}). In
a similar way, one can also represent classical open stochastic
systems, though we do not do so here. $\Box$
\end{example}

\subsection{Dissipative Networks}
\label{sec:qdiss-nw}

Dissipation properties for QFNs can be analyzed by considering how
the network is constructed in terms of the elementary constructs
(concatenation, series connection, direct interaction, LFT) and
knowledge of how dissipation \lq\lq{transforms}\rq\rq \  under
these constructs. This is the purpose of the following results.
The first  lemma considers concatenation and series connections.

 \begin{lemma}    \label{lemma:qdiss-transform}
Let $ {P}_1$ and $ {P}_2$ be given plants and $P_1 \wedge W_1$ and
$P_2 \wedge W_2$ are network structures to be specified. Assume
$P_1$ and $P_2$ are dissipative with respect to supply rates $r_{
{P}_1}( {W_1})$ and $r_{ {P}_2}( {W_2})$, storage functions $V_1$ and
$V_2$, and exosystem classes  $\ \mathscr{W}_1$ and $\
\mathscr{W}_2$ respectively. Then:
 \begin{enumerate}
 \item
 % 1
 If $P_1 \wedge W_1 = P_1 \boxplus W_1$ and $P_2 \wedge W_2 = P_2 \boxplus W_2$, then the network $P_1 \boxplus P_2$ is dissipative with respect to the network structure $(P_1 \boxplus P_2) \triangleleft  W$ and exosystem class $\mathscr{W}_1 \boxplus \mathscr{W}_2$; the storage function is $V=V_1 + V_2$ and the supply rate is
 \begin{equation}
 r_{P_1 \boxplus P_2} (W_1 \boxplus W_2) = r_{P_1}(W_1) + r_{P_2}(W_2) + \mathcal{G}_{P_1 \triangleleft W_1}(V_2) + \mathcal{G}_{P_2 \triangleleft  W_2}(V_1)  .
 \label{trans-concat}
 \end{equation}
 If $V_1$ commutes with $\mathscr{A}_{P_2}$ and $\mathscr{W}_{2}$ and if $V_2$ commutes with $\mathscr{A}_{P_1}$ and $\mathscr{W}_{1}$,   then
then the supply rate is simply the sum of the supply rates: $ r_{P_1 \boxplus P_2} (W_1 \boxplus W_2) = r_{P_1}(W_1) + r_{P_2}(W_2)$.

 %2

 \item
 If $P_1 \wedge W_1 = P_1 \triangleleft W_1$ and $P_2 \wedge W_2 = P_2 \triangleleft W_2$, then the network $P_2 \triangleleft P_1$ is dissipative with respect to the network structure $P_2 \triangleleft P_1 \triangleleft  W$ and exosystem class
 \begin{equation}
 \mathscr{W}= \{ W \ : \ P_2^\prime \triangleleft W \in \mathscr{W}_1 \ \mathrm{and} \ P_1 \triangleleft W \in \mathscr{W}_2 \},
 \label{trans-series-exo}
 \end{equation}
 where $P_2^\prime = P_1^{-1} \triangleleft P_2 \triangleleft P_1 = (S_1^\dagger S_2 S_1, S_1^\dagger(S_2-I)L_1+S_1^\dagger L_2, H_2+ \mathrm{Im} \{     L_2^\dagger (S_2+I)L_1-L_1^\dagger S_2L_1 \})$;
 the storage function is $V=V_1 + V_2$ and the supply rate is
 \begin{equation}
 r_{P_2 \triangleleft P_1} (W) = r_{P_1}(P_2^\prime \triangleleft W) + r_{P_2}(P_1 \triangleleft W)  .
  \label{trans-tri}
 \end{equation}

 \end{enumerate}
 \end{lemma}

\begin{proof}
 For part 1, we have
 \begin{eqnarray*}
 \mathcal{G}_{ (  {P}_1 \boxplus  {P}_2 ) \triangleleft ( {W}_1 \boxplus  {W}_2)}(V_1+V_2) &=&
  \mathcal{G}_{  ( {P}_1 \triangleleft   {W}_1) \boxplus ( {P}_2 \triangleleft   {W}_2)}(V_1+V_2)
  \\
  &=&  \mathcal{G}_{    {P}_1 \triangleleft  {W}_1 }(V_1) + \mathcal{G}_{    {P}_2 \triangleleft  {W}_2 }(V_2) + \mathcal{G}_{    {P}_1 \triangleleft   {W}_1 }(V_2) + \mathcal{G}_{    {P}_2 \triangleleft   {W}_2 }(V_1)
  \\
  & \leq &
  r_{ {P}_1}( {W}_1) + r_{ {P}_2}( {W}_2)
   + \mathcal{G}_{    {P}_1 \triangleleft   {W}_1 }(V_2) + \mathcal{G}_{    {P}_2 \triangleleft   {W}_2 }(V_1) ,
 \end{eqnarray*}
 so that the last line is a supply rate for the concatenated system.

 Similarly, for part 2 we have
 \begin{eqnarray*}
 \mathcal{G}_{  {P}_2 \triangleleft  {P}_1 \triangleleft  {W} } (V_1+V_2) &=&
  \mathcal{G}_{  {P}_2 \triangleleft  {P}_1 \triangleleft  {W} } (V_1) +  \mathcal{G}_{  {P}_2 \triangleleft  {P}_1 \triangleleft  {W} } (V_2)
  \\
  &=&   \mathcal{G}_{  {P}_1 \triangleleft ( {P}_2' \triangleleft  {W}) } (V_1) +  \mathcal{G}_{  {P}_2 \triangleleft ( {P}_1 \triangleleft  {W}) } (V_2)
  \\
  & \leq &
   r_{ {P}_1}( {P}_2'  \triangleleft  {W})+r_{ {P}_2}( {P}_1 \triangleleft    {W})
\end{eqnarray*}
 where $P_2^\prime = P_1^{-1} \triangleleft P_2 \triangleleft P_1 = (S_1^\dagger S_2 S_1, S_1^\dagger(S_2-I)L_1+S_1^\dagger L_2, H_2+ \mathrm{Im} \{     L_2^\dagger (S_2+I)L_1-L_1^\dagger S_2L_1 \})$ (\cite[Theorem 3.4]{GJ08}).
%$\Box$
\end{proof}

In the next lemma, we consider the dissipation properties of a  LFT feedback system in terms of series plant-exosystem network structures.

\begin{lemma}    \label{lemma:qdiss-LFT}
Let $P$ be a plant of the form (\ref{G-2x2}) that is dissipative with supply rate $r(W)$ and storage function $V$ with respect to the network structure $P \triangleleft W$ and exosystem class $\mathscr{W}$. Assume the LFT system $F(P)$ is well-defined (Figure \ref{fig:nw-elim1}, section \ref{sec:open-qfn}). Define a class
$$
F(\mathscr{W} ) = \{ (1,w,0)   \ : \
(1,w,0)  \boxplus ( 1, (I-S_{22})^{-1} (S_{21} w +L_2)  ,0   )
   \in \mathscr{W}
  \}
$$
Then $F(P)$ is dissipative with storage function $V$ and supply rate
\begin{equation}
r_{F(P)}( W) = r_P( (1,w,0)  \boxplus ( 1, (I-S_{22})^{-1} (S_{21} w +L_2)  ,0   ) )
\label{r-LFT}
\end{equation}
for $W=(1,w,0)\in F(\mathscr{W})$ relative to the network structure $F(P) \triangleleft W$.
\end{lemma}

\begin{proof}
Let $W=(1,w,0) \in F(\mathscr{W} )$, and consider the system $F(P) \triangleleft W$. By elimination of the internal signal in the feedback loop, we see that
$$
\tilde W = (1,w,0)  \boxplus ( 1, (I-S_{22})^{-1} (S_{21} w +L_2)  ,0   )
$$
is an admissible exosystem for $P$.  The result now follows from the assumed dissipation  property for $P$.
\end{proof}

The next lemma describes how series architectures may be used to modify the plant and supply rates.

\begin{lemma}    \label{lemma:qdiss-into-r}
Let $P$ and $Q$ be  systems for which the series connection $P \triangleleft Q$ is well defined, and assume
\begin{equation}
 {W} \in \mathscr{W} \ \ \textrm{implies} \ \
Q \triangleleft  {W}  \ \mathrm{and} \   Q^{-1}  \triangleleft  {W}  \in \mathscr{W}.
\label{no-Q}
\end{equation}
Then the plant  $ {P}$  is dissipative with storage function $V$ and supply rate $r_P( {W})$ with respect to $\mathscr{W}$ and series network architecture $P \triangleleft W$ if and only if the plant  $P \triangleleft Q$  is dissipative with storage function $V$ and supply rate $r_P(Q  \triangleleft  {W})$ with respect to $\mathscr{W}$ and series network architecture  $P \triangleleft Q \triangleleft W$.
\end{lemma}

\begin{proof}
The assertions follow from the relation
$$
\mathcal{G}_{P \triangleleft Q \triangleleft W}(V) = r_P( Q \triangleleft W).
$$
\end{proof}

As a consequence of Lemma \ref{lemma:qdiss-into-r},  the scattering matrix $ {S}$ for the plant  can be moved into the supply rate.

\begin{corollary}    \label{corol:qdiss-no-S}
Let $\mathscr{W}$ be a class of exosystems satisfying
\begin{equation}
 {W} \in \mathscr{W} \ \ \textrm{implies} \ \
( {S},0,0)  \triangleleft  {W}  \ \mathrm{and} \   ( {S}^\dagger,0,0) \triangleleft  {W}  \in \mathscr{W}.
\label{no-S-1}
\end{equation}
Then:
\begin{enumerate}
\item
The plant  $ {P}=( {S},  {L}, H)$  is dissipative with supply rate $r( {W})$ with respect to $\mathscr{W}$ series architecture  $P \triangleleft W$if and only if the plant  $ {P}'=( {1},  {L}, H)$  is dissipative with supply rate $r'( {W})=r(( {S}^\dagger,0,0) \triangleleft  {W})$ with respect to $\mathscr{W}$ and series architecture  $P' \triangleleft W$.
\item
The plant  $ {P}=( I,  {L}, H)$  is dissipative with supply rate $r( {W})$ with respect to $\mathscr{W}$ series architecture  $P \triangleleft W$if and only if the plant  $ {P}'=( S,  {L}, H)$  is dissipative with supply rate $r'( {W})=r(( {S},0,0) \triangleleft  {W})$ with respect to $\mathscr{W}$ and series architecture  $P' \triangleleft W$.
\end{enumerate}
\end{corollary}

The following example illustrates the use of some of the above results by considering the dissipation properties of a network analogous to \cite[Fig. 1 and Theorem 5, sec. 4]{WIL72}.

\begin{example}   \label{eg:willems-net}
Consider a network $N$ consisting of $n$ plants $P_1, \ldots, P_n$ interconnected via field channels and a static connection system $T$ (with complex numerical entries), Figure \ref{fig:qfn-willems}. In terms of the notation of this paper, $N=F(P)$, where $P= \tilde P \triangleleft (I \boxplus \tilde T)$, $\tilde P = Q_2 \triangleleft (P_1 \boxplus \cdots \boxplus P_n) \triangleleft Q_1$, and $\Tilde T= Q_4 \triangleleft T \triangleleft Q_3$. Here, $Q_1, Q_2, Q_3, Q_4$ are appropriately sized permutation matrices (static components that interchange signals).

 \begin{figure}[h]
\begin{center}

\setlength{\unitlength}{1579sp}%
\begingroup\makeatletter\ifx\SetFigFont\undefined%
\gdef\SetFigFont#1#2#3#4#5{%
  \reset@font\fontsize{#1}{#2pt}%
  \fontfamily{#3}\fontseries{#4}\fontshape{#5}%
  \selectfont}%
\fi\endgroup%
\begin{picture}(14744,6545)(2679,-8984)
\put(13951,-8911){\makebox(0,0)[lb]{\smash{{\SetFigFont{5}{6.0}{\familydefault}{\mddefault}{\updefault}{\color[rgb]{0,0,0}(b)}%
}}}}
\thicklines
{\color[rgb]{0,0,0}\put(3901,-3661){\vector(-1, 0){1200}}
}%
{\color[rgb]{0,0,0}\put(3901,-8161){\framebox(1800,1500){}}
}%
{\color[rgb]{0,0,0}\put(6901,-8161){\framebox(1800,5700){}}
}%
{\color[rgb]{0,0,0}\put(6901,-3661){\vector(-1, 0){1200}}
}%
{\color[rgb]{0,0,0}\put(5701,-2761){\vector( 1, 0){1200}}
}%
{\color[rgb]{0,0,0}\put(2701,-2761){\vector( 1, 0){1200}}
}%
{\color[rgb]{0,0,0}\put(6901,-7861){\vector(-1, 0){1200}}
}%
{\color[rgb]{0,0,0}\put(3901,-7861){\vector(-1, 0){1200}}
}%
{\color[rgb]{0,0,0}\put(2701,-6961){\vector( 1, 0){1200}}
}%
{\color[rgb]{0,0,0}\put(5701,-6961){\vector( 1, 0){1200}}
}%
{\color[rgb]{0,0,0}\put(11701,-4861){\framebox(1800,1500){}}
}%
{\color[rgb]{0,0,0}\put(11701,-3661){\vector(-1, 0){1200}}
}%
{\color[rgb]{0,0,0}\put(14701,-4861){\framebox(1500,750){}}
}%
{\color[rgb]{0,0,0}\put(17401,-4486){\vector(-1, 0){1200}}
}%
{\color[rgb]{0,0,0}\put(14701,-4486){\vector(-1, 0){1200}}
}%
{\color[rgb]{0,0,0}\put(11701,-4486){\line(-1, 0){1200}}
\put(10501,-4486){\line( 0,-1){1875}}
\put(10501,-6361){\line( 1, 0){6900}}
\put(17401,-6361){\line( 0, 1){1875}}
}%
{\color[rgb]{0,0,0}\put(17401,-3661){\vector(-1, 0){3900}}
}%
{\color[rgb]{0,0,0}\put(14701,-4111){\dashbox{225}(1500,750){}}
}%
{\color[rgb]{0,0,0}\put(11401,-5161){\dashbox{225}(5100,2100){}}
}%
\put(4576,-7561){\makebox(0,0)[lb]{\smash{{\SetFigFont{5}{6.0}{\familydefault}{\mddefault}{\updefault}{\color[rgb]{0,0,0}$P_n$}%
}}}}
\put(4576,-3361){\makebox(0,0)[lb]{\smash{{\SetFigFont{5}{6.0}{\familydefault}{\mddefault}{\updefault}{\color[rgb]{0,0,0}$P_1$}%
}}}}
\put(7726,-5461){\makebox(0,0)[lb]{\smash{{\SetFigFont{5}{6.0}{\familydefault}{\mddefault}{\updefault}{\color[rgb]{0,0,0}$T$}%
}}}}
\put(4801,-5461){\rotatebox{90.0}{\makebox(0,0)[lb]{\smash{{\SetFigFont{5}{6.0}{\familydefault}{\mddefault}{\updefault}{\color[rgb]{0,0,0}. . . .}%
}}}}}
\put(12376,-4186){\makebox(0,0)[lb]{\smash{{\SetFigFont{5}{6.0}{\familydefault}{\mddefault}{\updefault}{\color[rgb]{0,0,0}$\tilde P$}%
}}}}
\put(15151,-4561){\makebox(0,0)[lb]{\smash{{\SetFigFont{5}{6.0}{\familydefault}{\mddefault}{\updefault}{\color[rgb]{0,0,0}$\tilde T$}%
}}}}
\put(5401,-8836){\makebox(0,0)[lb]{\smash{{\SetFigFont{5}{6.0}{\familydefault}{\mddefault}{\updefault}{\color[rgb]{0,0,0}(a)}%
}}}}
{\color[rgb]{0,0,0}\put(3901,-3961){\framebox(1800,1500){}}
}%
\end{picture}%

\caption{(a) Quantum  network $N$ formed by  interconnecting systems with a static connection system $T$ ( \cite[Fig. 1]{WIL72}). (b) Equivalent representation of the network $N=F(\tilde P \triangleleft (I \boxplus \tilde T))$. }
\label{fig:qfn-willems}
\end{center}
\end{figure}
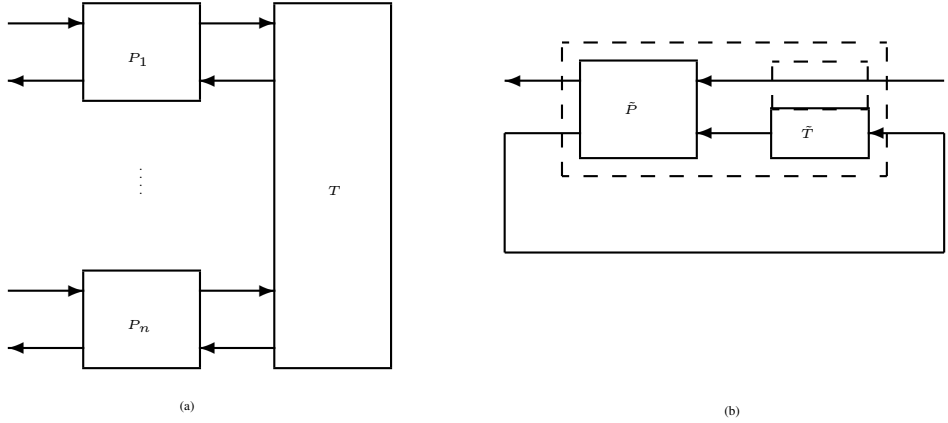

 We assume that the network $N$ of Figure \ref{fig:qfn-willems} is well-defined and that
for $j=1,\ldots, n$ let $P_j=(I, L_j ,H_j)$ be dissipative systems with supply rates $r_j(\tilde W_j)$ and storage functions $V_j$ with respect to the network structures  $P_j \triangleleft \tilde W_j$ and exosystem classes $\tilde {\mathscr{W}}_j$.  
%and that $\{w_j \}$ satisfy $[wj,w_k]=0$ and $[w_j,X_k]=0$ for all plant operators $X_k \in \mathscr{A}_{P_k}$.
Since the network $N=F(P)$ has $n$ input channels corresponding to the channels not involved in the interconnection, we may consider a series architecture $N \triangleleft W$, where $W=\boxplus_j W_j$ with appropriately sized exosystems of the form $W_j=(1,w_j,0)$.
Then
we can see that
with respect to the network structure $N \triangleleft W$ the network $N$ is dissipative with storage function $V=V_1+\ldots + V_n$ as follows.

First, by the first part of Lemma \ref{lemma:qdiss-transform}, we know that $\boxplus_j P_j$ is dissipative with storage function $V$ and supply rate
$$
r_{\boxplus_j P_j} ( \boxplus \tilde W_j) = \sum_j r_{P_j}(\tilde W_j)
$$
with respect to the network structure $(\boxplus_j P_j) \triangleleft (\boxplus_j \tilde W_j)$, where $\tilde W_j \in \tilde{\mathscr{W}}_j$.
Next,
we write
 $
 P=  R \triangleleft S
 $
 where
 $$
 R = Q_2 \triangleleft (\boxplus_j P_j ) = \boxplus_j R_j ,
 \ \
 S=Q_1 \triangleleft  (I \boxplus (Q_4 \triangleleft T \triangleleft  Q_3)),
 $$
 and $R_j = (I,  L_{\pi_2(j)}, 0)$. Here $\pi_2$ is the permutation corresponding to $Q_2$. Now $R$ is simply a re-arrangement of $\boxplus_jP_j$, and so it is dissipative with supply rate
 $$
 r_R( \tilde W) = \sum_j r_{P_{\pi_2(j)}}(\tilde W_j),
 $$
where $\tilde W=\boxplus \tilde W_j $, storage function $V$ and series architecture.

 Write $\hat W=\boxplus_j \hat W_j = \boxplus_j (I, \hat w_j, 0)$ and $\tilde W = S \triangleleft \hat W = (1, S\hat w,0)$.
   Then $P=R \triangleleft S$ is dissipative with respect to the series structure $P \triangleleft \hat W$ with supply rate
 $$
 r_P(\hat W) = r_R( \tilde W) = \sum_j r_{P_{\pi_2(j)}}(  \sum_k S_{jk} w_k  ).
 $$
  The supply rate for the network $N=F(P)$ now follows from Lemma \ref{lemma:qdiss-LFT}. The exosystem class can be determined from these calculations.
 $\Box$
\end{example}

\subsection{Uncertainty Modeling}
\label{sec:qdiss-uncert}

Because of its importance to questions of robustness, in this
subsection we briefly discuss how uncertainty can be modeled in
the framework of this paper. However, a detailed treatment of
robustness is beyond the scope of the present paper. The
plant-exosystem network architecture $P \wedge W$ illustrated in
Figure \ref{fig:pw} together with a specification of exosystem
class $\mathscr{W}$  provides a general scheme capable of
accommodating a wide range of uncertainty models.  Here for
definiteness we set $P \wedge W = P \triangleleft W$, and consider
a simple but common situation of parameter uncertainty.

Consider a plant $ {P}=( {I},  {L}, H)$, where
\begin{equation}
 {L}=(1+\epsilon)  {L}_0, \ \ H=H_0 + D .
\label{uncert-1}
\end{equation}
Here $\epsilon$ is a real parameter. Then using (\ref{series-dfn}) we can write
\begin{equation}
 {P}=  {P}_0 \triangleleft  {W} = ( {I},  {L}_0, H)  \triangleleft ( {I}, \epsilon  {L}_0, D) ,
\label{uncert-1a}
\end{equation}
which represents the plant $ {P}$ as a nominal system $ {P}_0$
with the uncertainty connected into it from an exosystem  $ {W}$
containing the uncertainty terms.  Note that $P$ and $W$ are not
independent systems, and in fact share variables.

The following example looks at the effect of parameter uncertainty
on the behavior of the damped oscillator (recall Example
\ref{eg:cavity-1}).

\begin{example}   \label{eg:qc-5}
(Parameter uncertainty) As an example of this type of uncertainty
modeling, consider  the plant of Example \ref{eg:cavity-1}, where
$\alpha = \sqrt{\gamma}$, and $\beta=0$ (damped oscillator).
Uncertainty arises from the parameter values $\gamma$ and
$\omega$, which  may not be known accurately. This type of
uncertainty is common in quantum optics, where $\gamma$ is a
measure of mirror  transmissivity and $\omega$ corresponds to a
detuning of the cavity relative to the input field. In the above
notation, let $L_0=\sqrt{\gamma_0}\, a$, $H_0= 0$ denote the
nominal parameters, defining the nominal plant $
{P}_0=(1,\sqrt{\gamma_0}\, a,  0 )$, a tuned cavity. Let the true
parameters be $L =(1+\epsilon)\sqrt{\gamma_0}\, a$, $H=\omega
a^\ast a$, a detuned cavity  $
{P}=(1,(1+\epsilon)\sqrt{\gamma_0}\, a,  \omega a^\ast a )$. Hence
the uncertain exosystem is  $ {W}=(1, \epsilon \sqrt{\gamma_0}\,
a, \omega a^\ast a)$ (so that $w=\epsilon \sqrt{\gamma_0}\, a$ and
$D=\omega a^\ast a$).

Then a straightforward calculation using (\ref{G-gen-def}) shows that
\begin{eqnarray*}
\mathcal{G}_{ {P}}(a^\ast a) &=& \mathcal{G}_{ {P}_0}(a^\ast a) - (\epsilon^2 +2\epsilon)\gamma_0 a^\ast a
\\ &=&
-(1+\epsilon)^2 a^\ast a \leq 0 .
\end{eqnarray*}
This means that the true system is always stable, regardless of
the uncertainty, though the decay rate does change. This of course
is expected of a passive physical system.

However, other aspects of system performance may be affected. For
instance, consider the quadratures $q=a+a^\ast$, $p=-i(a-a^\ast)$,
which are rotated by the detuning $\omega$. Indeed, if we focus on
the observable $q^2$, then the uncertain exosystem contribution to
the true  generator (recall (\ref{G-gen-def}))  is
$$
-i[ q^2, \omega a^\ast a] = \omega (qp+pq) ,
$$
which depends on the detuning parameter $\omega$. We see,
therefore, that a detailed robustness analysis with respect to
parameter uncertainty may involve  consideration of one or more
physical variables. $\Box$
\end{example}

\section{Feedback Control Design by Interconnection}
\label{sec:qcontrol}

\subsection{Control Design Formulation}
\label{sec:qcontrol-problem}

In the previous section we described how the behavior of plant is
influenced by exogenous influences (represented as exosystems)
using a quantum network framework. We now employ this same
framework to consider problems of control system design. The
network framework allows us to efficiently express a range of
control design problems. Indeed, in what follows we describe a
control design procedure that extends classical methods including
passivity-based control (PBC), \lq\lq{energy shaping}\rq\rq \
control, interconnection and damping assignment (IDA) PBC, and
\lq\lq{control as interconnection}\rq\rq, \cite{OS89},
\cite{VDS96}, \cite{OSMM01}, \cite{OSME02}, \cite{JW97},
\cite{JW98},  \cite{TW99}, \cite{TW02}, \cite{WT02}.

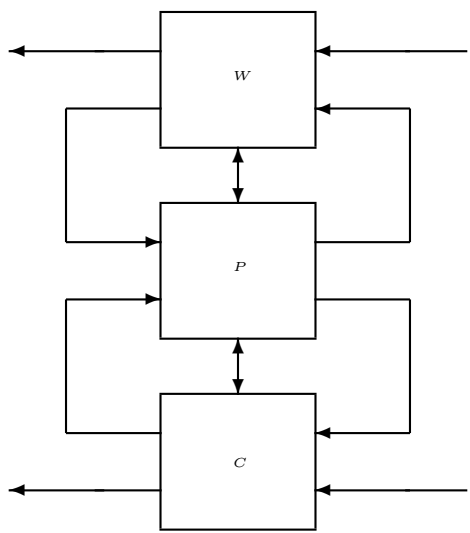
\begin{figure}[h]
\begin{center}

\setlength{\unitlength}{1579sp}%
\begingroup\makeatletter\ifx\SetFigFont\undefined%
\gdef\SetFigFont#1#2#3#4#5{%
  \reset@font\fontsize{#1}{#2pt}%
  \fontfamily{#3}\fontseries{#4}\fontshape{#5}%
  \selectfont}%
\fi\endgroup%
\begin{picture}(7244,8144)(1479,-9683)
\put(5026,-8686){\makebox(0,0)[lb]{\smash{{\SetFigFont{5}{6.0}{\familydefault}{\mddefault}{\updefault}{\color[rgb]{0,0,0}$C$}%
}}}}
\thicklines
{\color[rgb]{0,0,0}\put(7801,-2161){\vector(-1, 0){1500}}
}%
{\color[rgb]{0,0,0}\put(3001,-2161){\vector(-1, 0){1500}}
}%
{\color[rgb]{0,0,0}\put(2851,-2161){\line( 1, 0){1050}}
}%
{\color[rgb]{0,0,0}\put(7726,-2161){\line( 1, 0){975}}
}%
{\color[rgb]{0,0,0}\put(3901,-6661){\framebox(2400,2100){}}
}%
{\color[rgb]{0,0,0}\put(5101,-3661){\vector( 0, 1){  0}}
\put(5101,-3661){\vector( 0,-1){900}}
}%
{\color[rgb]{0,0,0}\put(3901,-3061){\line(-1, 0){1500}}
\put(2401,-3061){\line( 0,-1){2100}}
\put(2401,-5161){\vector( 1, 0){1500}}
}%
{\color[rgb]{0,0,0}\put(6301,-5161){\line( 1, 0){1500}}
\put(7801,-5161){\line( 0, 1){2100}}
\put(7801,-3061){\vector(-1, 0){1500}}
}%
{\color[rgb]{0,0,0}\put(5101,-6661){\vector( 0, 1){  0}}
\put(5101,-6661){\vector( 0,-1){900}}
}%
{\color[rgb]{0,0,0}\put(3901,-9661){\framebox(2400,2100){}}
}%
{\color[rgb]{0,0,0}\put(2851,-9061){\line( 1, 0){1050}}
}%
{\color[rgb]{0,0,0}\put(3001,-9061){\vector(-1, 0){1500}}
}%
{\color[rgb]{0,0,0}\put(7801,-9061){\vector(-1, 0){1500}}
}%
{\color[rgb]{0,0,0}\put(7726,-9061){\line( 1, 0){975}}
}%
{\color[rgb]{0,0,0}\put(3901,-8161){\line(-1, 0){1500}}
\put(2401,-8161){\line( 0, 1){2100}}
\put(2401,-6061){\vector( 1, 0){1500}}
}%
{\color[rgb]{0,0,0}\put(6301,-6061){\line( 1, 0){1500}}
\put(7801,-6061){\line( 0,-1){2100}}
\put(7801,-8161){\vector(-1, 0){1500}}
}%
\put(5026,-2611){\makebox(0,0)[lb]{\smash{{\SetFigFont{5}{6.0}{\familydefault}{\mddefault}{\updefault}{\color[rgb]{0,0,0}$W$}%
}}}}
\put(5026,-5611){\makebox(0,0)[lb]{\smash{{\SetFigFont{5}{6.0}{\familydefault}{\mddefault}{\updefault}{\color[rgb]{0,0,0}$P$}%
}}}}
{\color[rgb]{0,0,0}\put(3901,-3661){\framebox(2400,2100){}}
}%
\end{picture}%

\caption{The plant-controller-exosystem network $P \wedge C \wedge  W$.}
\label{fig:pwc}
\end{center}
\end{figure}

Let $ {P}$ be a plant which can be interconnected with a
controller $ {C}$ and an exosystem $W$  in a network $ {P} \wedge
{C} \wedge W$, Figure \ref{fig:pwc}, as described in section
\ref{sec:open-qfn}.  Given an exosystem class $\mathscr{W}_d$ we
denote by $\mathscr{C}$ a class of control systems for which such
a network is defined. Since the network parameters $ {P} \wedge
{C} \wedge W$ (together with the  inputs) determine the dynamical
evolution of the plant-controller-exosystem network, control
design can be thought of in terms of shaping the network dynamics
in a desired way by appropriate choice of controller $ {C}\in
\mathscr{C}$. Note that this framework is general enough to allow
us to consider choices for the controller that are quantum, or
classical, or a mixture of the two (cf. \cite{JNP08}).

 It is worth remarking that the distinction between plant and controller is not always clear-cut. Indeed, the plant may need to be engineered in an appropriate way to facilitate control. For example, one may desire   certain field couplings or direct interaction Hamiltonians to  be physically available---this clearly relates to the design of the plant.  This issue, of course, is not unknown in classical control engineering, and is evident in the examples below.
Also, if one is interested in the expected behavior of the network
for a range of variables $X$, then it may be appropriate to work
directly with the network  generator $\mathcal{G}_{ {P} \wedge {C}
\wedge W}(X)$, and use the form of this generator to determine the
controller $ {C}$, given the objectives. This approach is roughly
dual to a method based on master equations already in use, see,
e.g.  \cite{TW02}.

We specify the control objectives by encoding them in a
non-negative  observable $V_d \in  \mathscr{A}_{ {P}} \otimes
\mathscr{A}_{ {c}}$ (the parameters of the controllers $ {C} \in
\mathscr{C}$ are assumed to belong to $\mathscr{A}_{c}$), a supply
rate $r_d( {W})$, and a class of exosystems $\mathscr{W}_d$ for
which a network $( {P} \wedge  {C}) \wedge  {W}$ is well defined.
One then seeks to find, if possible,  a controller $ {C} \in
\mathscr{C}$ such that
\begin{equation}
\mathcal{G}_{  {P} \wedge  {C} \wedge  {W}}(V_d) -  r_d( {W}) \leq 0
\label{qc-diss-d}
\end{equation}
for all exosystem parameters $ {W} \in \mathscr{W}_d$. In other
words, one seeks a controller for which the closed loop system is
dissipative with storage function $V_d$, supply rate $r_d( {W})$,
and exosystem class $\mathscr{W}_d$. The exosystems are included
to facilitate robust control system design for situations where
uncertainty and disturbances are important.  The observable $V_d$
is something which on average should be small (such as regulation
errors), or tend to zero as time evolves. The supply rate may also
contain such \lq\lq{small}\rq\rq \ quantities, as well as terms
from the exosystems.  The supply rate need not be the natural
supply rate for the network---the inequality in the dissipation
inequality can be exploited to permit other choices.

\subsection{Controller Synthesis}
\label{sec:qcontrol-synth}

We shall now describe how standard problems of stabilization, regulation, and robust control fall within the scope of the controller synthesis framework formulated in the previous subsection. We begin with a general synthesis problem that abstracts stabilization and regulation, since they are closely related. These problems correspond to a choice $V_d$ of a non-negative observable whose expected value we wish to go to zero  as time approaches infinity. For definiteness, we choose $r_d( {W})=-c V_d$, where $c > 0$ is a suitable real number, and  $\mathscr{W}_d=\{ (\_,\_,0) \}$, which consists only of the trivial exosystem, so that $( {P} \wedge  {C} )\wedge  {W} = ( {P} \wedge  {C} )\boxplus  (\_,\_,0) =  {P} \wedge  {C}$.

\begin{theorem}   \label{thm:reg-stab}
(Stabilization/regulation)
If there exists a controller $ {C} \in \mathscr{C}$ and non-negative  observable $V_d \in  \mathscr{A}_{ {P}} \otimes \mathscr{A}_{c}$ such that the plant-controller network $ {P} \wedge  {C}$ satisfies
\begin{equation}
\mathcal{G}_{ {P} \wedge  {C} }(V_d) +   c V_d \leq 0
\label{qc-stab-1}
\end{equation}
for some real $c>0$, then $\langle V_d(t) \rangle \to 0$ as
$t \to \infty$ exponentially for any plant-controller state.  
\end{theorem}

This theorem follows from the stability results give in subsection
\ref{sec:qdiss-stability}.

Our next result is a general theorem concerning nonlinear quantum
$H^\infty$ robust control, which generalizes the linear quantum
results given in \cite{JNP08}, \cite{HM08}. We use  an exosystem
class $\mathscr{W}_{d}=\{  {W} \ : \  {W}=( {I},  {w},0) \ : \
{w} \ \mathrm{commutes\ with} \ \mathscr{A}_{ {P}} \}$ to describe
the \lq\lq{disturbance}\rq\rq\ inputs. The next  theorem is a
consequence of a slight extension of the Bounded Real Lemma
(Theorem \ref{thm:brl}) applied to the plant-controller network $
{P} \wedge  {C}$.

\begin{theorem}   \label{thm:hinfty}
($H^\infty$ control) If there exists a controller $ {C}\in
\mathscr{C}$ and a non-negative observable $V_d \in \mathscr{A}_{
{P}} \otimes \mathscr{A}_{ {c}}$ such that the plant-controller
network $ {P} \wedge  {C}$ satisfies
\begin{eqnarray}
  g^2 - {Z}^\dagger  {Z}   \geq  0 \
 \label{brl-Gamma-hinfty}
 \end{eqnarray}
 and
\begin{equation}
\mathcal{G}_{(  {P} \wedge  {C}) \wedge  {W}}(V_d)
- g^2  {w}^\dagger  {w}
  +
 ( {N}+ {Z} {w})^\dagger ( {N} + {Z} {w}) - \lambda \leq 0
\label{qc-hinfty-1}
\end{equation}
for some real $g > 0$, $\lambda \geq 0$, and all exosystem
parameters $ {W}  \in \mathscr{W}_d$, then the plant-controller
network $ {P} \wedge  {C}$ has gain $g$.
\end{theorem}

\subsection{Design Examples}
\label{sec:qcontrol-eg}

\begin{figure}[h]
\begin{center}

\setlength{\unitlength}{1579sp}%
\begingroup\makeatletter\ifx\SetFigFont\undefined%
\gdef\SetFigFont#1#2#3#4#5{%
  \reset@font\fontsize{#1}{#2pt}%
  \fontfamily{#3}\fontseries{#4}\fontshape{#5}%
  \selectfont}%
\fi\endgroup%
\begin{picture}(9344,2144)(2379,-6683)
\put(6826,-5461){\makebox(0,0)[lb]{\smash{{\SetFigFont{5}{6.0}{\familydefault}{\mddefault}{\updefault}{\color[rgb]{0,0,0}$B=\tilde A$}%
}}}}
\thicklines
{\color[rgb]{0,0,0}\put(7801,-5611){\vector(-1, 0){1500}}
}%
{\color[rgb]{0,0,0}\put(7801,-6661){\framebox(2400,2100){}}
}%
{\color[rgb]{0,0,0}\put(3901,-5611){\vector(-1, 0){1500}}
}%
{\color[rgb]{0,0,0}\put(11701,-5611){\vector(-1, 0){1500}}
}%
\put(5026,-5611){\makebox(0,0)[lb]{\smash{{\SetFigFont{5}{6.0}{\familydefault}{\mddefault}{\updefault}{\color[rgb]{0,0,0}$P$}%
}}}}
\put(8926,-5686){\makebox(0,0)[lb]{\smash{{\SetFigFont{5}{6.0}{\familydefault}{\mddefault}{\updefault}{\color[rgb]{0,0,0}$C$}%
}}}}
\put(3001,-5461){\makebox(0,0)[lb]{\smash{{\SetFigFont{5}{6.0}{\familydefault}{\mddefault}{\updefault}{\color[rgb]{0,0,0}$\tilde B$}%
}}}}
\put(10651,-5461){\makebox(0,0)[lb]{\smash{{\SetFigFont{5}{6.0}{\familydefault}{\mddefault}{\updefault}{\color[rgb]{0,0,0}$A$}%
}}}}
{\color[rgb]{0,0,0}\put(3901,-6661){\framebox(2400,2100){}}
}%
\end{picture}%

\caption{The plant-controller network   $P \triangleleft C$ for Examples \ref{eg:qc-1}, \ref{eg:qc-4}.}
\label{fig:pc-series}
\end{center}
\end{figure}
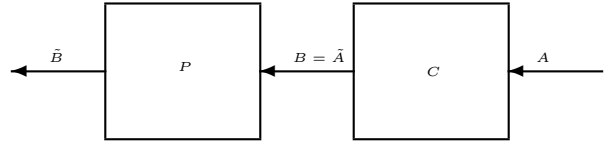

In this section we provide some simple examples to illustrate
several issues concerning feedback control design of quantum
systems. The  examples (Examples \ref{eg:qc-1}, \ref{eg:qc-4})
employ the series architecture shown in Figure
\ref{fig:pc-series}.

Our first design example is a regulation problem analogous to the
classical problem of designing a controller to maintain a given
value of capacitor charge in a RC circuit, \cite{OSMM01}.

\begin{example}   \label{eg:qc-1} (Regulation)
Consider an optical cavity
   $ {P}=(1,a,0)$   (a damped open harmonic oscillator,  Example \ref{eg:cavity-1}). If the input field is a vacuum, photons initially in the cavity will eventually leak out. Suppose our control objective is to maintain a given non-zero value for the steady state expected photon number.  Let's choose a value $\alpha$ for the desired steady-state value of $a$, which corresponds to a number $\alpha^\ast \alpha$ of photons (i.e., we want the cavity to be in a coherent state $\vert \alpha \rangle$ in the steady-state, Appendix \ref{sec:app-qho}).

Perhaps the simplest thing to do is to provide a source of fresh
photons that can be supplied to the cavity to replace those that are
lost. This might be achieved using a laser source or modulator
$C=(1,\nu,0)$ connected in series, as in Figure
\ref{fig:pc-series}. Here, $\nu$ is a complex number describing
the strength of the source, and is to be determined, if possible.

We set
$$
V_d = (a-\alpha)^\ast (a-\alpha) = a^\ast a -\alpha^\ast a - a^\ast \alpha + \alpha^\ast\alpha ,
$$
and for a positive real number $c$,
$$
r_d( {W})=-c V_d ,
$$
with $\mathscr{W}_d=\{ (\_,\_,0) \}$, which consists only of the
trivial exosystem, as in Theorem \ref{thm:reg-stab}. Note that the
expected value of $V_d$ in the state $\alpha$ is zero: $\langle
\alpha \vert V_d \vert \alpha \rangle=0$.

The design problem is to select  $\nu$, a complex number, such that
$$
\mathcal{G}_{  {P} \triangleleft   {C} } (V_d) + c V_d  \leq 0
$$
for suitable $c>0$.  Then from (\ref{lind-gen-series})
the LHS of this expression is
$$
-a^\ast a(1-c) +a(\frac{\alpha^\ast}{2}-\nu^\ast -c\alpha^\ast) +a^\ast( \frac{\alpha}{2}-\nu -c\alpha ) +\alpha^\ast \nu+\alpha\nu^\ast +c \alpha^\ast \alpha .
$$
If we set $c=1/2$, $\nu = -\alpha/2$, then this expression equals $-V_d/2$. Therefore
$$
\mathcal{G}_{  {P} \triangleleft   {C} } (V_d) \leq -V_d/2
$$
which implies that the expected value of $V_d(t)$ tends to zero as $t\to\infty$ (by Theorem \ref{thm:strict-passive-stability}), and the control objective is achieved (notes also that the expected value of $ {P} \triangleleft  {C} $ in the cavity coherent state $\vert \alpha \rangle$ is zero: $\langle \alpha \vert \mathcal{G}_{  {P} \triangleleft  {C} } (V_d)    \vert \alpha \rangle=0$).

The effect of the controller is to place the cavity input field in a coherent state $\vert \nu \rangle$. It is well known (e.g. \cite{GZ00}) that this is equivalent to adding a Hamiltonian term to the cavity model and setting the input to vacuum. In the notation of this paper, this follows from
 (\ref{lind-gen-concat}) and  (\ref{lind-gen-series}): $\mathcal{G}_{  {P} \triangleleft  {C} } (X)    = \mathcal{G}_{  P \boxplus \tilde {C} } (X)$, where $\tilde C= (\_,\_, -i(\nu a^\ast - \nu^\ast a))$. This is illustrated in Figure \ref{fig:pc-direct}. $\Box$

 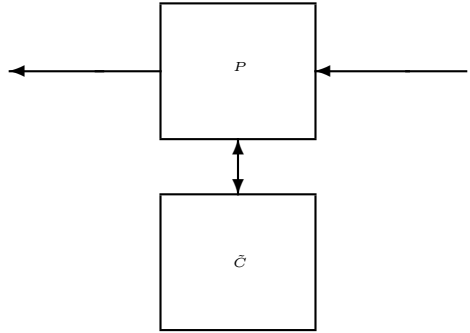
\begin{figure}[h]
\begin{center}

\setlength{\unitlength}{1579sp}%
\begingroup\makeatletter\ifx\SetFigFont\undefined%
\gdef\SetFigFont#1#2#3#4#5{%
  \reset@font\fontsize{#1}{#2pt}%
  \fontfamily{#3}\fontseries{#4}\fontshape{#5}%
  \selectfont}%
\fi\endgroup%
\begin{picture}(7244,5144)(1479,-9683)
\put(5026,-8686){\makebox(0,0)[lb]{\smash{{\SetFigFont{5}{6.0}{\familydefault}{\mddefault}{\updefault}{\color[rgb]{0,0,0}$\tilde{C}$}%
}}}}
\thicklines
{\color[rgb]{0,0,0}\put(5101,-6661){\vector( 0, 1){  0}}
\put(5101,-6661){\vector( 0,-1){900}}
}%
{\color[rgb]{0,0,0}\put(3901,-9661){\framebox(2400,2100){}}
}%
{\color[rgb]{0,0,0}\put(2851,-5611){\line( 1, 0){1050}}
}%
{\color[rgb]{0,0,0}\put(7801,-5611){\vector(-1, 0){1500}}
}%
{\color[rgb]{0,0,0}\put(7726,-5611){\line( 1, 0){975}}
}%
{\color[rgb]{0,0,0}\put(3001,-5611){\vector(-1, 0){1500}}
}%
\put(5026,-5611){\makebox(0,0)[lb]{\smash{{\SetFigFont{5}{6.0}{\familydefault}{\mddefault}{\updefault}{\color[rgb]{0,0,0}$P$}%
}}}}
{\color[rgb]{0,0,0}\put(3901,-6661){\framebox(2400,2100){}}
}%
\end{picture}%

\caption{Alternate representation of the plant-controller  network $P \triangleleft C$ in the form  $P \boxplus \tilde C  $ for Example \ref{eg:qc-1}.}
\label{fig:pc-direct}
\end{center}
\end{figure}

\end{example}

The next example shows that care must be exercised when attempting
to use classical control design methods.   Furthermore,
one must take account of quantum noise and  the fact that physical
quantities do not in general commute.

\begin{example}   \label{eg:qc-4}
(Stabilization) Suppose we wish to stabilize the marginally stable
system $P=(1, a+a^\ast, 0)$, a special case of the open
oscillators of Example \ref{eg:cavity-1}, using an approach
analogous to a  standard method from classical control theory for
stabilizing Hamiltonian systems, \cite[sec. 4.1]{VDS96}.

Consider the series plant-controller network of Figure
\ref{fig:pc-series}. From Example \ref{eg:cavity-1}, we know that
this system is passive; if $C=(1,u,0)$ then
\begin{equation}
\mathcal{G}_{P \triangleleft C}(V_0)  = \mathscr{L}_u(V_0) +   u^\ast Z + Z^\ast u   +1,
\label{pc-gen-marginal}
\end{equation}
where $Z=  a^\ast -a$ and $V_0=a^\ast a$ (recall (\ref{qdiss-general-r-1}), (\ref{r0-oscillator}) with $W=C$).
The classical Hamiltonian stabilization procedure suggests that we set
\begin{equation}
u = -k Z
\label{stab-1}
\end{equation}
for some non-negative gain $k$. In order to implement this
feedback, the controller $C$ needs to have access to the variable
$Z$. However, $Z$ is not available in the output signal $\tilde
B$, since by the output relation (\ref{out-Lambda}) $d\tilde
B=Ldt+dB$, where $L=a+a^\ast$.

We suppose that the plant can be re-engineered to have a second
field channel which contains $Z$. Specifically, we consider the
augmented system $P \boxplus C$, where $C=(1, -kZ,0)$.  The desired variable is now available in the second output of the augmented system, and so we can form the series connection $P \triangleleft C = (1,  (1+k)a +(1-k)a^\ast, - ik(a^2-(a^\ast)^2) )$, Figure \ref{fig:p-marginal}. However, an examination of the dynamics of the quadratures $q=a+a*$ and $p=-i(a-a^\ast)$ shows that the feedback system $P \triangleleft C$ is marginally stable and not asymptotically stable  for all $k \geq 0$ (the feedback system has poles at $0$ and $-4k$). Physically, neither of the field couplings $L=a+a^\ast =q$ and $-kZ =k(a-a^\ast)=ikp$ are sufficient for strict passivity and hence asymptotic stability. 

A preferable stabilization scheme would be to replace $C$ by $\tilde C = (1, ka, 0)$. Then the  re-engineered systems $P \boxplus \tilde C$ and $P \triangleleft \tilde C$ are both strictly passive and asymptotically stable.

Note that the ability to engineer field couplings is of fundamental importance here (see \cite{NJD08} for
general results concerning physical realization in the context of
linear quantum systems).  

\begin{figure}[h]
\begin{center}

\setlength{\unitlength}{1579sp}%
\begingroup\makeatletter\ifx\SetFigFont\undefined%
\gdef\SetFigFont#1#2#3#4#5{%
  \reset@font\fontsize{#1}{#2pt}%
  \fontfamily{#3}\fontseries{#4}\fontshape{#5}%
  \selectfont}%
\fi\endgroup%
\begin{picture}(6344,4844)(2379,-9383)
\put(7051,-7261){\makebox(0,0)[lb]{\smash{{\SetFigFont{5}{6.0}{\familydefault}{\mddefault}{\updefault}{\color[rgb]{0,0,0}$A$}%
}}}}
\thicklines
{\color[rgb]{0,0,0}\put(3901,-5611){\vector(-1, 0){1500}}
}%
{\color[rgb]{0,0,0}\put(3901,-8461){\framebox(2400,3900){}}
}%
{\color[rgb]{0,0,0}\multiput(3901,-6511)(436.36364,0.00000){6}{\line( 1, 0){218.182}}
}%
{\color[rgb]{0,0,0}\put(7801,-7411){\vector(-1, 0){1500}}
}%
{\color[rgb]{0,0,0}\put(3901,-7411){\line(-1, 0){1500}}
\put(2401,-7411){\line( 0,-1){1950}}
\put(2401,-9361){\line( 1, 0){6300}}
\put(8701,-9361){\line( 0, 1){3750}}
\put(8701,-5611){\line(-1, 0){1200}}
}%
\put(5026,-5611){\makebox(0,0)[lb]{\smash{{\SetFigFont{5}{6.0}{\familydefault}{\mddefault}{\updefault}{\color[rgb]{0,0,0}$P$}%
}}}}
\put(3001,-5461){\makebox(0,0)[lb]{\smash{{\SetFigFont{5}{6.0}{\familydefault}{\mddefault}{\updefault}{\color[rgb]{0,0,0}$\tilde B$}%
}}}}
\put(6976,-5461){\makebox(0,0)[lb]{\smash{{\SetFigFont{5}{6.0}{\familydefault}{\mddefault}{\updefault}{\color[rgb]{0,0,0}$B=\tilde A$}%
}}}}
\put(5026,-7486){\makebox(0,0)[lb]{\smash{{\SetFigFont{5}{6.0}{\familydefault}{\mddefault}{\updefault}{\color[rgb]{0,0,0}$C$}%
}}}}
{\color[rgb]{0,0,0}\put(7801,-5611){\vector(-1, 0){1500}}
}%
\end{picture}%

\caption{Alternative representation of the plant-controller network  $P \triangleleft C$ for Example \ref{eg:qc-4}.}
\label{fig:p-marginal}
\end{center}
\end{figure}
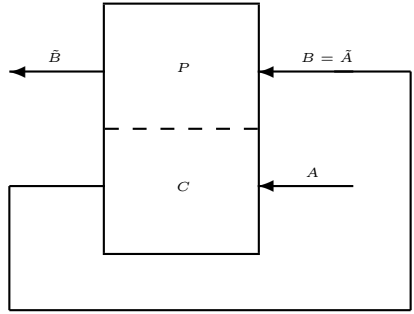

$\Box$
\end{example}

\section{Conclusions}
\label{sec:conclusion}

In this paper we have extended J.C.~Willems' theory of dissipative
systems to the quantum domain.  The quantum systems we considered
are open quantum models, and with the aid of recently developed
methods for describing quantum feedback networks, we have shown
how to describe external influences as arising from interactions
with exosystems. The fundamental dissipation property was
expressed in these terms. We presented an infinitesimal
characterization of the dissipation property, which  generalizes
the well-known Positive Real and Bounded Real Lemmas. We also
showed how to implement Willems'  \lq\lq{control by
interconnection}\rq\rq \   for open quantum   systems using
quantum network representations.

We believe that the results in this paper provide useful methods
for the analysis and design of quantum dissipative systems, and
indeed networks of such systems. The quantum network based results
we have presented   are quite general and powerful, and merit
further development. The network paradigm is particularly
important if quantum technology is to move from the device and
small system level to a more complex system level such as is being
contemplated, for example,  in the quantum computing community.

\appendix

\subsection{Two Level Atom (Qubit)}
\label{sec:app-qbit}

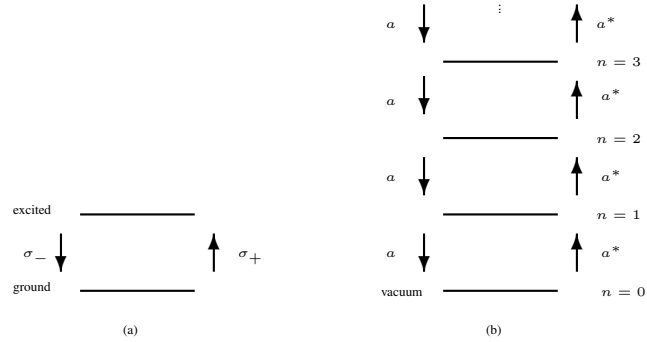
\begin{figure}[h]
\begin{center}

\setlength{\unitlength}{1579sp}%
\begingroup\makeatletter\ifx\SetFigFont\undefined%
\gdef\SetFigFont#1#2#3#4#5{%
  \reset@font\fontsize{#1}{#2pt}%
  \fontfamily{#3}\fontseries{#4}\fontshape{#5}%
  \selectfont}%
\fi\endgroup%
\begin{picture}(9255,5270)(1036,-10109)
\put(10201,-5836){\makebox(0,0)[lb]{\smash{{\SetFigFont{5}{6.0}{\familydefault}{\mddefault}{\updefault}{\color[rgb]{0,0,0}$n=3$}%
}}}}
\thicklines
{\color[rgb]{0,0,0}\put(2101,-8161){\line( 1, 0){1800}}
}%
{\color[rgb]{0,0,0}\put(7801,-9361){\line( 1, 0){1800}}
}%
{\color[rgb]{0,0,0}\put(1801,-8461){\vector( 0,-1){600}}
}%
{\color[rgb]{0,0,0}\put(7501,-8461){\vector( 0,-1){600}}
}%
{\color[rgb]{0,0,0}\put(7801,-8161){\line( 1, 0){1800}}
}%
{\color[rgb]{0,0,0}\put(7801,-6961){\line( 1, 0){1800}}
}%
{\color[rgb]{0,0,0}\put(7801,-5761){\line( 1, 0){1800}}
}%
{\color[rgb]{0,0,0}\put(7501,-7261){\vector( 0,-1){600}}
}%
{\color[rgb]{0,0,0}\put(7501,-5986){\vector( 0,-1){600}}
}%
{\color[rgb]{0,0,0}\put(9901,-9061){\vector( 0, 1){600}}
}%
{\color[rgb]{0,0,0}\put(9901,-7861){\vector( 0, 1){600}}
}%
{\color[rgb]{0,0,0}\put(9901,-6661){\vector( 0, 1){600}}
}%
{\color[rgb]{0,0,0}\put(9901,-5461){\vector( 0, 1){600}}
}%
{\color[rgb]{0,0,0}\put(7501,-4861){\vector( 0,-1){600}}
}%
{\color[rgb]{0,0,0}\put(4201,-9061){\vector( 0, 1){600}}
}%
\put(1201,-8836){\makebox(0,0)[lb]{\smash{{\SetFigFont{5}{6.0}{\familydefault}{\mddefault}{\updefault}{\color[rgb]{0,0,0}$\sigma_-$}%
}}}}
\put(1051,-8161){\makebox(0,0)[lb]{\smash{{\SetFigFont{5}{6.0}{\familydefault}{\mddefault}{\updefault}{\color[rgb]{0,0,0}excited}%
}}}}
\put(1051,-9361){\makebox(0,0)[lb]{\smash{{\SetFigFont{5}{6.0}{\familydefault}{\mddefault}{\updefault}{\color[rgb]{0,0,0}ground}%
}}}}
\put(2776,-10036){\makebox(0,0)[lb]{\smash{{\SetFigFont{5}{6.0}{\familydefault}{\mddefault}{\updefault}{\color[rgb]{0,0,0}(a)}%
}}}}
\put(6826,-9436){\makebox(0,0)[lb]{\smash{{\SetFigFont{5}{6.0}{\familydefault}{\mddefault}{\updefault}{\color[rgb]{0,0,0}vacuum}%
}}}}
\put(6901,-6436){\makebox(0,0)[lb]{\smash{{\SetFigFont{5}{6.0}{\familydefault}{\mddefault}{\updefault}{\color[rgb]{0,0,0}$a$}%
}}}}
\put(10276,-8836){\makebox(0,0)[lb]{\smash{{\SetFigFont{5}{6.0}{\familydefault}{\mddefault}{\updefault}{\color[rgb]{0,0,0}$a^\ast$}%
}}}}
\put(10276,-7636){\makebox(0,0)[lb]{\smash{{\SetFigFont{5}{6.0}{\familydefault}{\mddefault}{\updefault}{\color[rgb]{0,0,0}$a^\ast$}%
}}}}
\put(10276,-6361){\makebox(0,0)[lb]{\smash{{\SetFigFont{5}{6.0}{\familydefault}{\mddefault}{\updefault}{\color[rgb]{0,0,0}$a^\ast$}%
}}}}
\put(6901,-5236){\makebox(0,0)[lb]{\smash{{\SetFigFont{5}{6.0}{\familydefault}{\mddefault}{\updefault}{\color[rgb]{0,0,0}$a$}%
}}}}
\put(10201,-5236){\makebox(0,0)[lb]{\smash{{\SetFigFont{5}{6.0}{\familydefault}{\mddefault}{\updefault}{\color[rgb]{0,0,0}$a^\ast$}%
}}}}
\put(6901,-7636){\makebox(0,0)[lb]{\smash{{\SetFigFont{5}{6.0}{\familydefault}{\mddefault}{\updefault}{\color[rgb]{0,0,0}$a$}%
}}}}
\put(6901,-8836){\makebox(0,0)[lb]{\smash{{\SetFigFont{5}{6.0}{\familydefault}{\mddefault}{\updefault}{\color[rgb]{0,0,0}$a$}%
}}}}
\put(4576,-8836){\makebox(0,0)[lb]{\smash{{\SetFigFont{5}{6.0}{\familydefault}{\mddefault}{\updefault}{\color[rgb]{0,0,0}$\sigma_+$}%
}}}}
\put(8476,-10036){\makebox(0,0)[lb]{\smash{{\SetFigFont{5}{6.0}{\familydefault}{\mddefault}{\updefault}{\color[rgb]{0,0,0}(b)}%
}}}}
\put(8701,-5011){\rotatebox{90.0}{\makebox(0,0)[lb]{\smash{{\SetFigFont{5}{6.0}{\rmdefault}{\mddefault}{\updefault}{\color[rgb]{0,0,0}...}%
}}}}}
\put(10276,-9436){\makebox(0,0)[lb]{\smash{{\SetFigFont{5}{6.0}{\familydefault}{\mddefault}{\updefault}{\color[rgb]{0,0,0}$n=0$}%
}}}}
\put(10201,-8236){\makebox(0,0)[lb]{\smash{{\SetFigFont{5}{6.0}{\familydefault}{\mddefault}{\updefault}{\color[rgb]{0,0,0}$n=1$}%
}}}}
\put(10201,-7036){\makebox(0,0)[lb]{\smash{{\SetFigFont{5}{6.0}{\familydefault}{\mddefault}{\updefault}{\color[rgb]{0,0,0}$n=2$}%
}}}}
{\color[rgb]{0,0,0}\put(2101,-9361){\line( 1, 0){1800}}
}%
\end{picture}%

\caption{Energy level diagrams. (a) Two-level atom (qbit). (b) Harmonic oscillator.}
\label{fig:levels}
\end{center}
\end{figure}

The simplest quantum system has two energy levels and is often
used to model ground and excited states of atoms. Since the advent
of quantum computing, this system is also known as the qubit, the
unit of quantum information. The two level atom is illustrated in
Figure \ref{fig:levels} (a), showing the action of the raising
$\sigma_+$ and lowering $\sigma_-$ operators. The Hilbert space
for this system is $\mathsf{H}=\mathbf{C}^2$, the two-dimensional
complex vector space. The physical variable space $\mathscr{A}$
for this system is spanned by the Pauli matrices \cite[sec.
2.1.3]{NC00}, \cite[sec. 9.1.1]{GZ00}:
$$
\sigma_0 = I= \left(  \begin{array}{cc}  1 & 0 \\ 0 & 1\end{array}  \right), \ \
\sigma_x = I= \left(  \begin{array}{cc}  0 & 1 \\ 1 & 0 \end{array}  \right), \ \
\sigma_y = I= \left(  \begin{array}{cc}  0 & -i \\ i & 0 \end{array}  \right),  \\
\sigma_z = I= \left(  \begin{array}{cc}  1 & 0 \\ 0 & - 1\end{array}  \right).
$$
The raising and lowering operators are defined by
$\sigma_{\pm}=\frac{1}{2}(\sigma_x \pm i\sigma_y)$. The basic
commutation relations are $[\sigma_x, \sigma_y]=2i \sigma_z$,
$[\sigma_y, \sigma_z]=2i \sigma_x$, and $[\sigma_z, \sigma_x]=2i
\sigma_y$. The energy levels correspond to the eigenvalues of
$\sigma_z$.

\subsection{Quantum Harmonic Oscillator}
\label{sec:app-qho}

The quantum harmonic oscillator is one of the most important
examples because of its tractability and application to modeling,
\cite[Box 7.2]{NC00}, \cite[sec. 10.6]{EM98}, \cite[sec.
4.1]{GZ00}. Models for the optical cavity and boson fields are
based on the quantum harmonic oscillator. The quantum harmonic
oscillator is illustrated in Figure \ref{fig:levels} (b), which
shows infinite ladder of energy levels and the action of the
creation $a^\ast$ and annihilation $a$ operators. The Hilbert
space for the quantum harmonic oscillator is
$\mathsf{H}=L^2(\mathbf{R}, \mathbf{C})$, the vector space of
square integrable functions defined on the real line. The physical
variable space $\mathscr{A}$  for this system is defined in terms
of the annihilation operator $a$, with $a^\ast$ the adjoint of
$a$, and the {\em canonical commutation relations} $[a,a^\ast]=1$.
The action of the annihilation operator may be expressed as
$$
(a \psi)(x) = x \psi(x) -i \frac{d \psi}{dx}(x)
$$
on a domain of functions (vectors) $\psi$ in $\mathsf{H}$.  The
eigenvalues of $a^\ast a$ are the numbers $0,1,2, \ldots$ (number
of quanta), with corresponding eigenvectors denoted $\psi_n$
($n=0,1,2,\ldots$) called {\em number states}. We have $a \psi_n =
\sqrt{n}\, \psi_{n-1}$ and $a^\ast \psi_n = \sqrt{n+1}\,
\psi_{n+1}$.
  For a complex number $\alpha$, a {\em coherent state} is defined by
  $$
  \vert \alpha \rangle = \exp( -\frac{1}{2} \vert \alpha \vert^2 ) \sum_{n=0}^{\infty} \frac{\alpha^n}{\sqrt{n!}} \psi_n
  $$
  (Dirac notation),
and satisfies the eigenvalue relation $a \vert \alpha \rangle = \alpha \vert \alpha \rangle$.

\subsection{Operator Orderings}
\label{sec:app-order}

In this appendix we review some definitions and results concerning operator ordering.

Let $A$ and $B$ be self-adjoint operators on a Hilbert space
$\mathsf{H}$. Then by definition   $A \geq 0$ means $\langle \psi,
A \psi \rangle \geq 0$ for all vectors $\psi  \in \mathsf{H}$.
Using this, we say $A \geq B$ to mean $A - B \geq 0$.

Now fix $A$ and $B$ self-adjoint, and $C$ is arbitrary. Assume
\begin{equation}
w^\ast A w \leq B + w^\ast C + C^\ast w
  \label{op-1}
  \end{equation}
for all operators $w$ acting on $\mathsf{H}$. Then we claim that
$A \leq 0$.

To verify this claim, suppose by contradiction there exists
$\psi_0  \in \mathsf{H}$ such that
    \begin{equation}
  \langle \psi_0, A \psi_0 \rangle  > 0.
  \label{op-2}
  \end{equation}
Now set $w=\alpha I$, where $\alpha$ is an arbitrary real number.
Then (\ref{op-1}) implies
    \begin{equation}
  \alpha^2 \langle \psi_0, A \psi_0 \rangle  \leq \langle \psi_0, B \psi_0 \rangle + \alpha  \langle \psi_0,  (C+C^\ast ) \psi_0 \rangle  .
  \label{op-3}
  \end{equation}
Since $\alpha$ is arbitrary, this contradicts (\ref{op-2}),
establishing the claim.

Now fix arbitrary operators $C$ and $D$. We can use a similar
argument to show that if
     \begin{equation}
  w^\ast C +  C^\ast w  \leq D
  \label{op-4}
  \end{equation}
for all operators $w$ acting on $\mathsf{H}$, then $C=0$.

{\bf Acknowledgement.}  The authors wish to thank Guofeng Zhang for his helpful comments.

%%%%%%%%%%%%%%%%%%%%%%

\bibliographystyle{plain}
%\bibliography{mjbib2004}

\end{document}